\documentclass[12pt,a4paper]{article}%
\usepackage{amsfonts}
\usepackage{amsmath,amsthm,amssymb,natbib,verbatim,upref}
\usepackage{indentfirst,setspace,graphicx,rotating,fullpage,epsfig,float,chngpage,subfigure}
\usepackage{color}
\usepackage{amsmath}
\usepackage{amssymb}
\usepackage{graphicx}
\providecommand{\U}[1]{\protect\rule{.1in}{.1in}}
\newtheorem{theorem}{Theorem}
\newtheorem{proposition}{Proposition}
\newtheorem{assumption}{Assumption}
\newtheorem{lemma}{Lemma}
\newtheorem{corollary}{Corollary}

\theoremstyle{definition}

\newtheorem{example}{Example}
\newtheorem{excont}{Example}

\begin{document}

\title{Capacity Games with Supply Function Competition}
\author{Edward Anderson\thanks{University of Sydney Business School, Sydney, NSW 2006, Australia;
        edward.anderson@sydney. edu.au}
\and    Bo Chen\thanks{Warwick Business School, University of Warwick, Coventry, CV4 7AL, UK;
        b.chen@warwick.ac.uk}
\and    Lusheng Shao\thanks{Corresponding author. Faculty of Business and Economics, University of
        Melbourne, Melbourne, VIC 3010, Australia; lusheng.shao@unimelb.edu.au}
        }
\date{May 24, 2019}

\maketitle

\begin{abstract}
This paper studies a setting in which multiple suppliers compete for a buyer's
procurement business. The buyer faces uncertain demand and there is a requirement
to reserve capacity in advance of knowing the demand. Each supplier has costs that
are two dimensional, with some costs incurred before demand is realized in order to
reserve capacity and some costs incurred after demand is realized at the time of
delivery. A distinctive feature of our model is that the marginal costs may
not be constants, and this naturally leads us to a supply function competition
framework in which each supplier offers a schedule of prices and quantities.
We treat this problem as an example of a general class of capacity games and show that,
when the optimal supply chain profit is submodular, in
equilibrium the buyer makes a reservation choice that maximizes the overall
supply chain profit, each supplier makes a profit equal to their marginal
contribution to the supply chain, and the buyer takes the remaining profit. We
further prove that this submodularity property holds under two commonly
studied settings: (1) there are only two suppliers; and (2) in the case of
more than two suppliers, the marginal two-dimension costs of each supplier are non-decreasing and constant, respectively.



\end{abstract}

\section{Introduction}
\label{sec-intro}

When demand is uncertain, the characteristics of the supply chain and the contract arrangements will determine
investment decisions. In capital-intensive industries, such as petrochemical, electronics and semiconductors,
manufacturers need to invest heavily in building capacity and the lead times are long \citep{Kleindorfer2003},
so that it remains a critical challenge to incentivize and manage capacity investment.
When manufacturers are competing against each other and a buying
firm may switch to a different manufacturer, the problems are intensified. If the financial risks from investment
are carried by the firms who build capacity, it is difficult to find good solutions as there is no certainty on
how much capacity will be required in the future \citep{WuErkoc2005}. The result may be that capacity is
only built for the part of the demand that can be more or less guaranteed: this conservative approach will
reduce the buyer's ability to meet customers' demand. From a supply chain perspective, it is usually preferable
for buyers to reserve capacity from suppliers in advance, with a payment made to the suppliers associated with this
capacity reservation. Then the buyer also shares the risk that demand is low and not all the purchased capacity is required.

We model capacity reservation in a supply option framework. In the first
stage, before knowing the actual demand, a buyer reserves a certain amount of
capacity by paying a \textit{reservation price}. In the second stage, after
discovering the actual demand, the buyer asks for supply up to the
lesser of the reserved amount and the observed demand. At this stage, the
buyer pays an \textit{execution price} only for the amount of capacity that is
used. The underlying assumption of this model is that suppliers have to
install capacity, or place orders with their own suppliers, before demand materializes
because of the lead times involved.

A distinctive feature of our model is that marginal costs (of capacity and production) may depend on the volume,
which is often the case in practical situations. For example, capacity investment often involves a one-off setup
cost \citep{Luss82, van-Mieghem03}, there may be (dis)economies of scale in production \citep{Haldi&Whitcomb67,
HaTonZha11}, or in cases where a supplier manages a portfolio of facilities with heterogeneous technologies the
overall cost is far from being linear (especially in the electricity industry). In our model, each supplier has a
total reservation cost (incurred before demand realization), which is an arbitrary increasing function of the capacity
reserved, and also a total execution cost (incurred after demand realization), which
is an arbitrary increasing function of the amount actually supplied. This
framework includes constant marginal costs as a special case, so our model is
flexible enough to capture many practical settings in terms of cost modelling.
We believe this is an important theoretical advancement as most literature on supply chain contracts assumes constant
marginal costs \citep{Cachon03}. The linear treatment may help simplify the analysis and develop high-level intuitions
especially for competitive models, but may not align with real-world practices.

With constant marginal costs, it is plausible to focus on simple contract
forms, but with general cost functions, more sophisticated contract
formats may be worthwhile, such as bids which specify a schedule of different
prices for different quantities (i.e., a supply function). This type of supply
function bid often occurs in practice through the application of some form of
quantity discount contract.

In this paper, we consider a supply chain with a single buyer and multiple
suppliers, who supply a homogeneous item to the buyer. The suppliers compete by
offering price functions for both reservation and execution. Given the
suppliers' bids, the buyer first reserves capacity before knowing the actual
demand, and then decides how much capacity to use after observing the demand.
The buyer may choose not to enter into a contract with a specific supplier in
which case no costs are incurred by that supplier. Competition with function bids fits well with the situation
where a buyer does not stipulate the specific bidding format in its Request for Quote (RFQ), so
that suppliers can bid in whatever format they like.

We suppose that the buyer faces a random demand, with a known distribution.
There are $n$ suppliers, and each has a total reservation cost, which is expressed
as a general increasing function of the amount reserved, as well as a total execution
cost which is a general increasing function of the amount supplied. We model
this game in a Stackelberg framework, where the suppliers are leaders and the
buyer is a follower. Each supplier has complete information about the buyer's
demand distribution and all supplier costs, but the buyer may not know the
suppliers' costs. This assumption has been widely used in supplier competition models
and is appropriate for industries where the operating environments are more transparent and/or
the production technologies are more mature such as electricity, electronics and semiconductors
\citep{Wu2005, MartinzeSimchi2009, AndersonChenShao2017}.

\begin{figure}[ptbh]
\centering
\includegraphics[width=0.9\linewidth]{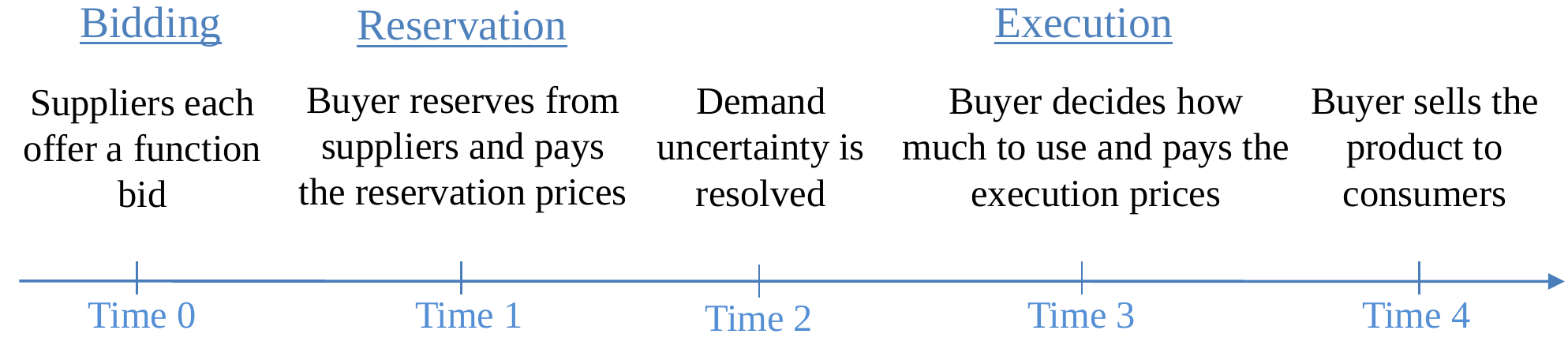}\caption{The
timeline}%
\label{fig:timeline}%
\end{figure}

The sequence of events is depicted in Figure~\ref{fig:timeline}.
First each supplier, with the aim of maximizing their own expected profits,
offers a function bid consisting of a (marginal) reservation price function
and a (marginal) execution price function to the buyer. Second, prior to
knowing the actual demand, the buyer determines how much to reserve from each
supplier and pays the reservation price. After demand is observed, the buyer
chooses what capacity to use and pays the execution price for the amount of
capacity that is used. If the demand exceeds the total amount of reserved
capacity, there will be lost sales. Finally, the buyer sells the product to
the consumer market at the retail price $\rho$. We will call this \emph{the capacity
game with function bids}. This game can be considered as a special case of a general capacity game setting,
in which a first stage choice by the buyer implies a constraint
on the decisions to be made by the buyer at a second stage when demand information becomes known. Thus,
our discussion will be carried out for the general capacity game, which in turn provides a theoretical
foundation for the capacity game with function bids.

The framework we consider also applies to capacity mechanisms in electricity markets.
In this context the buyer and customers are replaced by a set of consumers and a system
operator who aims to maximize system welfare. Problems of this form are usually formulated
with a value of lost load (VoLL). In the case where demand is stochastic but independent of price,
the system operator's problem, of minimizing the expected cost of procurement less
VoLL for demand not met, becomes equivalent to maximizing VoLL times (demand met) minus
the cost of procurement, which is the same as the buyer's problem in our supply chain setting when
we set the retail price to be $\textrm{VoLL}$.

It is not uncommon to have a supply chain where multiple suppliers act as
Stackelberg leaders and a buyer must select amongst the available suppliers on
the basis of contracts that the suppliers offer. This framework will lead to
suppliers setting as high a price as they can while still being chosen. The
consequence is that each supplier makes a bid that leads to the buyer
being indifferent between choosing or not choosing the supplier.
From this it can be shown that the profit available to each supplier is simply its
contribution to the overall supply chain profit, i.e., the difference in the
supply chain profits when it is chosen and when it is not.

A major contribution of this paper is to study submodularity in this setting.
We will demonstrate that a natural sufficient condition for a well-behaved
equilibrium is that the supply chain profits are submodular in terms of the
set of suppliers available. This property is equivalent to a requirement that the profit available to a supplier
cannot increase if a new supplier is added to those available. We are therefore led to considering whether or not
the supply chain profits achieved are submodular. This is not always the case,
but can be established in commonly studied settings. Specifically, the property holds
if either: (i) there are only two suppliers; or (ii)
each supplier's marginal reservation cost is non-decreasing and marginal execution cost is constant.

A second contribution of this paper is to study how suppliers compete with
each other by using supply functions rather than scalar prices. In other
words, the strategy space of each supplier is extended to a function space,
and as a result, the suppliers have more flexibility in their bidding
decisions. On the other hand, the buyer can determine how much to reserve from
each supplier (i.e., pick any point from a supply function) rather than being
restricted to a fixed quantity-payment contract. Our aim is to develop an
understanding of the players' strategic behavior in such a market setting. Previous models have focused
on competition problems when suppliers are restricted to a simple strategy space
(i.e., scalar decision variables, including prices, quantities, lead times,
quality, or specific contract types). To the best of our knowledge, this paper is among the first to study supply
function competition in a capacity reservation setting.


We find that given knowledge of the other suppliers' bids, it is optimal for each
supplier to set execution prices at execution costs and make profits only
through the buyer's reservation payment. Our findings also show that, under
some reasonable assumptions, there exists a continuum of equilibria in which
the suppliers charge costs only for the execution prices but add a margin to
the reservation costs. Despite the multiplicity of equilibrium bidding
strategies, the equilibrium outcome is essentially unique: in equilibrium the
buyer makes a reservation choice that maximizes the overall supply chain
profit, each supplier makes a profit that is equal to their marginal
contribution to the supply chain, and the buyer takes the remaining profit.

The rest of this paper is organized as follows. After a review of the relevant
literature in Section~\ref{sec-lit}, we present our general capacity game model
in Section~\ref{section-setup}, and study the best response and
equilibrium strategies for suppliers in Sections~\ref{sec:equilibrium}.
Then we show how these results apply to the more
specific case we have introduced above in Section~\ref{section-cap-game-functions}.
We discuss in Section~\ref{sec:submodularity} two commonly studied settings and show that
the supply chain optimal profit is submodular in a broad class of capacity games with function
bids. Finally, we conclude in Section~\ref{section-conclusion}. All technical proofs are presented in the
appendix.

\section{Literature review}
\label{sec-lit}

Supply options have been extensively studied in the operations management literature
\citep[see, e.g.,][]{Yehuda2002, Burnetas2005, Wudss2005, Fu2010, Secomandi2012}. An initial step is to
investigate a buyer's optimal purchasing decision given a fixed set of supply options \citep[see, e.g.,][]{martinez2005}.
As an extension of this, several papers examine option contract design problems in a Stackelberg game between a buyer and
a supplier with a focus on the interaction between option markets and spot markets \citep[see, e.g.,][]{Wu2002, Pei2011}.
Further extensions have been made to consider supplier competition in an option market \citep{Wu2005, MartinzeSimchi2009,
AndersonChenShao2017}. \cite{Wu2005} study a commodity procurement problem where an industrial buyer purchases supply
options from a small set of suppliers, while having access to an open spot market. No reservation cost at the suppliers is
considered, and the buyer's downward-sloping demand (arising from utilization maximization) is determined by the spot
price and the execution prices. Such a demand modelling approach allows the buyer to rank suppliers by using a single price
index combining the reservation and execution price of supply options, which leads to a Bertrand type of equilibrium for the
suppliers. We study a similar situation but consider suppliers having a two-dimensional cost and a newsvendor type buyer
facing uncertain demand.

The closest papers to ours within this literature are \citet{MartinzeSimchi2009} (hereafter, ``MS'') and \cite{AndersonChenShao2017}
(hereafter, ``ACS''). MS studies a setting where marginal costs are constants and each supplier chooses a reservation price and an execution price for their \textit{limitless} capacity. They show that there may be efficiency loss in equilibrium, which is up to $25\%$ of the overall supply chain optimal profit. Our paper differs from MS by considering general cost functions and allowing suppliers to submit function bids. By enlarging the strategy space of suppliers, we find a relatively clean and intuitive result as discussed earlier. In particular, when the supply chain optimal profit is submodular, there is no efficiency loss in equilibrium and each supplier makes a profit equal to their marginal contributions. We further show that this condition is not restrictive as it holds under some commonly studied settings. Comparing with MS, we also find that allowing suppliers to offer a function bid makes the buyer worse off, thereby highlighting the significant impact of the strategy space on equilibrium outcomes.

ACS studies a similar problem where each supplier owns a block of capacity with the same size and the buyer must reserve all
of a block or none. Our general framework of capacity games covers the model by ACS as a special case since we can recover ACS by restricting the buyer's capacity choice from each supplier to be either zero or the block size. In this respect, our paper generalizes ACS by studying a situation where the buyer can reserve any amount from each supplier and suppliers can charge different prices for different quantities.

This paper studies a situation where suppliers compete with each other by offering a function bid, and this resembles what is
studied in the supply function equilibrium (SFE) literature \citep{klemperer1989, anderson2002, anderson2008, Johari2011}.
However, the game type is different. Our model is a Stackelberg game with multiple leaders (i.e., suppliers) and one follower (i.e., buyer), and the buyer strategically responds to the supplier offers. In contrast, the SFE literature focuses on Nash games that aim to study the strategic interaction between multiple firms, and this literature does not involve a profit-maximizing buyer. In other words, we examine the buyer's optimization problem explicitly; while in the SFE literature, the buyer's problem is to choose a clearing price to equate the total supply with the demand, and each supplier's best response is characterized by a differential equation. In addition, our capacity games involve the buyer making a two-stage decision and each supplier submits a two-dimensional bid consisting of a reservation price function and execution price function, which is not considered in the SFE literature.

This work is also related to the multi-unit auction literature. Problems of decentralized resource allocation \citep{Kalagnanam2004}
are central to auctions. A subset of this literature examines the efficiency and profit allocation of a given type of auction, for example, share auction \citep{wilson1979}, menu auction \citep{bernheim1986}, split award auction \citep{Anton1989, Anton1992}, discriminatory price auction \citep{Menezes1995}, and uniform price auction \citep{Bresky2013205}. Those papers generally assume that the total purchase amount of a buyer is exogenously given, while in our model it is endogenously determined by the buyer based on the supplier bids. Furthermore, we study capacity reservation games where the buyer's capacity reservation problem is a two-stage stochastic program, which is new to this literature.

Finally, the equilibrium profit allocation obtained in this paper is of Vickrey-Clarke-Groves (VCG) style, so it is instructive to
discuss the key point of departure from VCG models \citep{Vickrey1961, Rothkopf1990}. Those models study profit allocation from a
mechanism design perspective: the VCG profit allocation is an outcome that arises from the payment rule. In essence the VCG
approach selects bids that maximize system welfare, and pays according to each bidder's contribution (i.e., the difference between
system welfare with and without that bidder). With this setup the equilibrium result is for bidders to bid truthfully and the
auction result is efficient. The model we propose uses the same bid selection approach, but pays as bid. It is then an equilibrium
(in the case with submodularity) for the bidders to mark up by the exact VCG amounts. We still get an efficient auction result, but
without truthful bids. This is a different approach and has some advantages over the usual VCG mechanism. \citet{Hobbs_et_al2000}
and \citet{Rothkopf07} amongst others have discussed some of the difficulties of using a VCG auction in practice. Many of these
problems are resolved with pay-as-bid prices. For example, in the auction format we discuss, there is not the same incentive
for the buyer to encourage bids that will not be accepted with the aim of reducing the payments to the successful bidders.
Note that a submodularity condition has also been discussed in the VCG framework by \citet{AusMil06}, who show that this condition
reduces the problems ordinarily associated with VCG.
Within the mechanism design literature, several papers study multi-dimensional bidding environments in an electricity context,
where the focus is on designing scoring and payment rules that incentivize bidders to truthfully reveal their costs
\citep{BusOre94, ChaWil02, Schummer2003}.

\section{Model setup}
\label{section-setup}

The problem we study is an example of a broader class of capacity games, in
which the suppliers offer bids, then a buyer selects a capacity amount to buy
from each supplier, then depending on the outcome of a random variable (the
demand) in the second stage, the choice of amounts to buy is made by the buyer (with
amounts constrained by the capacity already bought). The revenue to the buyer
is a function of demand and the amounts bought. We are not considering risk
aversion, so we can use the expected payoff as the objective for the buyer and
the suppliers. Supplier $i$ faces costs that are the sum of a capacity cost
and an execution cost. We will establish our results in this general framework
and then apply this to the model described in the introduction section.

The buyer makes capacity decisions $\mathbf{t}=(t_{1},\ldots,t_{n})$, where
capacity $t_{i}$ is restricted to lie in a set $T_{i}\subseteq \mathbb{R}_+$, and hence
$\mathbf{t}\in\mathbf{T}=T_{1}\times T_{2}\times \cdots \times T_{n}$. We assume
that each $T_{i}$ is compact containing $0$ and is
specified as part of the capacity game. Typically, $T_{i}$ is a closed interval
$[0, \bar{d}_i]$ or $T_{i}=\{0,W_{i}\}$, which corresponds to the case where the buyer either does not use
this supplier, or uses the full amount $W_{i}$ of the supplier.

The capacity payment to be made by the buyer to each supplier can depend on
the complete set of capacity decisions $\mathbf{t}$, which may occur in practice, for example, when the suppliers have some
exclusivity clauses. Note that, we are making our model as general as possible, although in the specific
applications of our model considered in Section~\ref{section-cap-game-functions}, the capacity payment made to a
supplier depends only on the buyer's choice from that supplier. In the second stage,
after demand $D\in \mathcal{D}$ becomes known, there is a set of amounts (execution
quantities) $\mathbf{s}\in\mathbf{T}$ chosen by the buyer, where $\mathbf{s}$ depends on the observed demand $D$
and satisfies constraints imposed by the capacity decisions of the first stage. Thus we have $\mathbf{s}\in X(\mathbf{t})$
for some set $X(\mathbf{t})\subseteq \mathbf{T}$. We require that if $t_{i}=0$ then $s_{i}=0$. The case that is
most natural is to allow any selection at the second stage that is no more than
the capacity purchase at the first stage, so that $\mathbf{s}\leq\mathbf{t}$
and $X(\mathbf{t})=\{\mathbf{s}:s_{i}\leq t_{i},i\in N\}$. In general, however, we only assume
$X(\mathbf{t})$ is compact. Then supplier $i$ decides on two payment functions $R_{i}(\cdot)$ and $P_{i}(\cdot)$
that the buyer will be offered, where $R_i(\mathbf{t})$ is the payment made to supplier $i$ when the buyer makes a
capacity reservation of $\mathbf{t}$, and $P_i(\mathbf{s})$ is the payment made to supplier $i$ given execution
amounts $\mathbf{s}$. In the formulation of the game there may be
restrictions on the bids allowed, which we capture by specifying an allowed set
of functions $A_{i}$, $i\in N$, and restrict the supplier's choice to
$R_{i}(\cdot)\in A_{i}$ and $P_{i}(\cdot)\in A_{i}$.
Since the buyer can decide not to include supplier $i$, in which case no payment is
made, the feasible bids $A_{i}$ have the property that for all $R_{i}(\cdot)\in A_{i}$,
$R_{i}(\mathbf{t})=0$ whenever $t_{i}=0$ (similarly for $P_{i}(\cdot)$). For maximal generality of our model, we
specify only the minimal further restrictions on $A_{i}$ at the end of this section.

Besides payments made by the buyer to the suppliers, we also have costs
$E_{i}(\mathbf{t)}$ and $C_{i}(\mathbf{s})$ incurred by supplier $i$ that may
depend on the complete set of capacities purchased and amounts executed.
This occurs when competing players can partially collaborate (e.g., they share warehouse facilities or transport links).
However, in the
capacity game, a supplier's costs may be independent of other suppliers' quantities, as in our specific applications
considered in Section~\ref{section-cap-game-functions}. We assume that these costs are zero if supplier $i$ is not used
at either stage and hence $E_{i}(\mathbf{t})=0$ when $t_{i}=0$, and $C_{i}(\mathbf{s})=0$ when
$s_{i}=0$.

Finally, we have the expected revenue $V(D,\mathbf{s)}$ to the buyer from an
external source, which depends on the demand $D$ and the execution amounts
$\mathbf{s}$. Rather than include a restriction that the total execution
amounts are no more than the demand (i.e., $\sum_{i\in N} s_{i}\leq D$), we will assume that
this is achieved through setting appropriate values for the revenue function
$V$.

\begin{example}\label{exp:model-illustration}
We consider an example that illustrates the range of modelling applications
for the general capacity game. Consider a buyer facing uncertain demand: $50$
with probability $0.5$ and $100$ with probability $0.5$. There are two
suppliers with identical cost structures who produce products that are fully
substitutable. We take $T_{1}=T_{2}=[0,100]$. Each supplier has costs of
reservation for an amount $t$ of capacity given by $t^{2}/150$ with an additional
fixed cost of \$20 associated with building a transport link. This fixed
cost is split between any suppliers who have capacity reserved, so it is
\$20 if just one supplier is used, and otherwise \$10 each. Thus the
reservation costs are
\[
E_{i}(\mathbf{t}) = \left\{
                      \begin{array}{ll}
                        {t_{i}^{2}}/{150}+20-10\chi (t_{j}),\ j\neq i, & \hbox{if $t_{i}>0$;} \\
                        0, & \hbox{if $t_{i}=0$;}
                      \end{array}
                    \right.
\]
where the indicator function $\chi (t_{j})=1$ if $t_{j}>0$, and $0$
otherwise. Execution costs are \$1 per unit so that  $C_{i}(\mathbf{s})=s_{i}$. Customers
pay an amount \$3 per unit and in addition there are fixed costs at the buyer of $\$20$
for each product that apply only if that product is supplied. Thus,
$V(D,\mathbf{s})=3\min(D,s_{1}+s_{2})-20\chi (s_{1})-20\chi (s_{2})$.
We have the standard form for $X(t)$ so there is a restriction that $s_{i}\leq t_{i}$. \ $\Box$
\end{example}

Since execution amounts will depend on $D$, we can write a policy for the buyer
as $(\mathbf{t},\mathbf{s}(\cdot))$, where $\mathbf{t} \in \mathbf{T}$ and $\mathbf{s}(\cdot)$ is a function from the
demand set $\cal{D}$ to $\mathbf{T}$, taking the value $\mathbf{s}(D)$ at $D$. For simplicity, we will write
$\mathbf{s}$ for $\mathbf{s}(\cdot)$ hereafter. The total expected buyer profit with this policy
choice is
\[
\Pi_{\mathcal{B}}(\mathbf{t},\mathbf{s})=\mathbb{E}[V(D,\mathbf{s}(D))
 -\sum_{i\in N}(P_{i}(\mathbf{s}(D))+R_{i}(\mathbf{t}))],
\]
given a set of supplier bids $\mathcal{B}=\{(P_{i}(\cdot), R_{i}(\cdot)) \in A_i: i\in N\}$.
The buyer's problem is to maximize her expected profit by choosing an optimal reservation
choice $\mathbf{t}$ and a set of execution amounts $\mathbf{s}(D)$ for each possible demand $D \in \cal{D}$, i.e., to solve
\begin{equation}\label{eq-buyer-formulation}
\max\left\{  \Pi_{\mathcal{B}}(\mathbf{t},\mathbf{s}):\mathbf{t}\in
\mathbf{T},\mathbf{s}(D)\in X(\mathbf{t}) \mbox{ for all } D\in \mathcal{D}\right\}.
\end{equation}
With these bids and buyer choice $(\mathbf{t},\mathbf{s})$, supplier $i$ has a profit of
\begin{equation}\label{eq:supplier-profit}
\pi_{i}(\mathbf{t},\mathbf{s}) = \mathbb{E}\left[  R_{i}
(\mathbf{t})-E_{i}(\mathbf{t})+P_{i}(\mathbf{s}(D))-C_{i}(\mathbf{s}(D))\right],\ i\in N.
\end{equation}
The supply chain profit is
\begin{align}\label{eqn:channel_profit}
\Pi_{\mathcal{C}}(\mathbf{t},\mathbf{s})  &  =\Pi_{\mathcal{B}}%
(\mathbf{t},\mathbf{s})+\sum_{i\in N}\pi_{i}(\mathbf{t},\mathbf{s}) \nonumber\\
&  =\mathbb{E}[V(D,\mathbf{s}(D))-\sum_{i\in N}(C_{i}(\mathbf{s}(D))+E_{i}(\mathbf{t}))],
\end{align}
which is independent of the bids made.

We write $I(\mathbf{t})=\{i:t_{i}>0\}$ for the support of a vector $\mathbf{t}$. We use
$\Pi_{\mathcal{C}}^{\ast}(S)$, $S\subseteq N$, to denote the optimal supply chain profit
when the capacity reserved is restricted to be zero outside the set $S$. Thus
\begin{equation}\label{eq:supplychain_prob}
\Pi_{\mathcal{C}}^{\ast}(S)=\max\left\{  \Pi_{\mathcal{C}}(\mathbf{t},
\mathbf{s}): \mathbf{t}\in\mathbf{T},\mathbf{s}(D)\in X(\mathbf{t}) \mbox{ for all } D\in \mathcal{D},
I(\mathbf{t})\subseteq S\right\}.
\end{equation}
Similarly, given a set of supplier bids $\mathcal{B}=\{(P_{i}(\cdot),
R_{i}(\cdot))\in A_i: i\in N\}$, we use $\Pi_{\mathcal{B}}^{\ast}(S)$,
$S\subseteq N$, to denote the optimal buyer profit when the capacity reserved is
restricted to be zero outside the set $S$. Thus
\[
\Pi_{\mathcal{B}}^{\ast}(S)=\max\left\{  \Pi_{\mathcal{B}}(\mathbf{t},
\mathbf{s}):\mathbf{t}\in\mathbf{T}, \mathbf{s}(D)\in X(\mathbf{t}) \mbox{ for all } D\in \mathcal{D},
I(\mathbf{t})\subseteq S\right\}.
\]
The maximizations above are taken over $\mathbf{t}\in \mathbf{T}$ (a vector) and $\mathbf{s}(\cdot)\in X(\mathbf{t})$ (a function).
To ensure that the maxima exist and are attained, we make the following (additional) assumptions: (a) the set-valued function
$X(\mathbf{t})$ is upper hemi-continuous; (b) the revenue function $V(\mathbf{t},\mathbf{s})$ is upper semi-continuous in both
arguments; and (c) the set $A_i$ of possible bids is such that $P_i(\cdot), R_i(\cdot)\in A_i$ are lower semi-continuous.

\section{Best response and equilibrium behavior}
\label{sec:equilibrium}

We now look at each supplier's best response problem. As a Stackelberg leader,
each supplier is able to anticipate the buyer's reservation choice provided
that the competitors' bids are observed. Since the buyer's optimization
problem is embedded in the suppliers' best responses, we need to specify the buyer's choice
when there are different options with the same expected value to the buyer. We say that
supplier $i$ has \emph{preferred status} when a solution with $t_{i}>0$ is chosen by the buyer even if other
options give the same value.

\begin{theorem}[Best Response]\label{thm:best-resp}
For any fixed supplier $i\in N$, we have the following statements for the supplier's best response:
\begin{enumerate}
\item[(a)] Given bids $\{(P_{j}(\cdot),R_{j}(\cdot)): j\in N, j\neq i\}$,
the profit for supplier $i$ is no more than
\[
Z_{i}=\Pi_{\mathcal{B}_{i}}^{\ast}(N)-\Pi_{\mathcal{B}_{i}}^{\ast}(N\backslash\{i\}),
\]
where $\mathcal{B}_{i}=\{(P_{j}(\cdot),R_{j}(\cdot)): j\in N, j\neq i\}\cup\{(C_{i}(\cdot),E_{i}(\cdot))\}$.
\item[(b)] If $Z_{i}>0$ and supplier $i$ has preferred status, then the profit $Z_{i}$ is
achieved by the offer $(\bar{P}_{i}(\cdot), \bar{R}_{i}(\cdot))$ defined by:
\begin{align*}
\bar{P}_{i}(\mathbf{s})  &  =C_{i}(\mathbf{s})\text{, and}\\
\bar{R}_{i}(\mathbf{t})  &  =E_{i}(\mathbf{t})+Z_{i}\text{ when }t_{i}>0, \bar{R}_{i}(\mathbf{t})=0\text{ when }t_{i}=0.
\end{align*}
\item[(c)] If $Z_{i}>0$, then for any $\epsilon>0$, an offer of $\left(\bar{P}_i(\cdot), \bar{R}_i ^{(\epsilon)}(\cdot)\right)$
will achieve within $\epsilon$ of the maximum possible supplier profit, where $\bar{R}_i^{(\epsilon)}(\mathbf{t})
=\bar{R}_i (\mathbf{t})- \epsilon$.
\end{enumerate}
\end{theorem}

Theorem~\ref{thm:best-resp} shows the maximum profit for a supplier $i$, which is the increase in profit available
to the buyer from including the bids of supplier $i$ when these are made at cost. Moreover, when optimizing for the
supplier, it is sufficient to consider supplier bids that are at cost for the execution component and make money only
from the reservation payments. However, we should note that the expected profit to the supplier is unaffected by parts
of the function bids that are never selected by the buyer whatever demand occurs. The consequence is that there are
a continuum of other best response function offers available.

Now let us consider the equilibrium strategies for suppliers.
In general, the optimal solution to the buyer's maximization problem \eqref{eq-buyer-formulation}
is not unique at an equilibrium as we shall show later. Since our focus is on supplier competition,
and each supplier's payoff depends on the buyer's reservation choice, we need a tie-breaking rule to pin
down the buyer's optimal choice in case of multiple solutions. The suppliers
have an interest in raising prices to the point where the buyer is about to
drop them from consideration. Therefore, a tie-breaking assumption is critical,
otherwise, we can have a difficulty in defining optimal behavior for the
suppliers, as a type of $\epsilon$-optimality could occur when a
supplier sets his price just below a benchmark value at
which the buyer no longer wishes to select the supplier. Consequently, we make the following assumption.

\begin{assumption}[Tie-Breaking Rule]\label{assum-SCopt}
In case of multiple optimal solutions, the buyer will select the one that gives the largest supply chain profit.
\end{assumption}

This assumption is in line with that for the classic Bertrand competition model, where the firm with the lowest cost
wins when all the firms charge the same price. The rationale of this assumption is as follows: if the buyer selects
an alternative that gives a lower supply chain profit, then keeping the other suppliers' bids unchanged, any supplier
in the supply chain optimal set (with respect to the buyer's equally good choices) will have incentives to raise his
price by an amount that is smaller than the difference between the supply chain optimal profit and the one corresponding
to the buyer's choice. This would make both the buyer and the supplier better off.

We need to rule out cases where two suppliers are identical and the optimal supply chain solution can use either one or
the other, so we introduce the following assumption.

\begin{assumption}[Uniqueness of Support]\label{assum-unique}
All the supply chain optimal capacity choices use the same set of suppliers.
\end{assumption}

We will write $(\mathbf{t}_{S}^{\ast},\mathbf{s}_{S}^{\ast})$ to denote an optimal solution to the supply chain optimization
problem \eqref{eq:supplychain_prob}. So Assumption~\ref{assum-unique} amounts to a statement that
the support $I(\mathbf{t}_{N}^{\ast})$ of the supply chain optimal capacity choices $\mathbf{t}_{N}^{\ast}$ is unique.
Let us identify the suppliers who make a contribution to the optimal supply chain
profit by defining
\begin{align*}
N^{\ast}(\mathcal{C}) &  =\{j\in N:\Pi_{\mathcal{C}}^{\ast}(N)>\Pi_{\mathcal{C}}^{\ast}(N\backslash\{j\})\},\\
N_{-i}^{\ast}(\mathcal{C})
 &  =\{j\in N:\Pi_{\mathcal{C}}^{\ast}(N\backslash\{i\})>\Pi_{\mathcal{C}}^{\ast}(N\backslash\{i,j\})\}.
\end{align*}
It is easy to observe that, under Assumption~\ref{assum-unique}, we have
\begin{equation}\label{eq:support_tNstar}
I(\mathbf{t}_{N}^{\ast})=N^{\ast}(\mathcal{C}).
\end{equation}

To establish an equilibrium, we will need to make use of submodularity of
the optimal supply chain profit as a set function of the suppliers available, where the submodularity property states that,
for any $i,j\in S\subseteq N$ and $i\neq j$,
\begin{equation}\label{eq-sub}
{\Pi}_{\mathcal{C}}^{\ast}(S)-{\Pi}_{\mathcal{C}}^{\ast}(S\setminus\{i\})
\leq{\Pi}_{\mathcal{C}}^{\ast}(S\setminus\{j\})-{\Pi}_{\mathcal{C}}^{\ast}(S\setminus\{i,j\}).
\end{equation}
Inequalities \eqref{eq-sub} are satisfied for a large class of problems of interest, as we shall
discuss in detail in Section~\ref{sec:submodularity}. It is convenient to make this property
an assumption:
\begin{assumption}[Submodularity]\label{ass:submodularity}
The supply chain optimal profit ${\Pi}_{\mathcal{C}}^{\ast}(S)$ is submodular as a function of the set $S\subseteq N$ of
available suppliers.
\end{assumption}

This submodularity assumption implies two properties of the set function ${\Pi}_{\mathcal{C}}^{\ast}(\cdot)$,
which are established in the following two lemmas.

\begin{lemma}\label{le-sub1}
Under the Submodularity Assumption, the following inequality holds for any $S\subseteq N$:
\begin{equation}\label{eq-sub-2}
{\Pi}_{\mathcal{C}}^{\ast}(N)-{\Pi}_{\mathcal{C}}^{\ast}(S)\geq\sum_{i\in N\setminus S}
\left({\Pi}_{\mathcal{C}}^{\ast}(N)-{\Pi}_{\mathcal{C}}^{\ast}(N\setminus\{i\})\right).
\end{equation}
\end{lemma}

The following result shows that all the suppliers in $N^{\ast}(\mathcal{C})\setminus\{i\}$
will still be included in the optimal supply chain choice when supplier $i$ is unavailable.

\begin{lemma}\label{le-sub2}
Under the Submodularity Assumption, for any $i\in N$, we have
$N^{\ast}(\mathcal{C})\setminus\{i\}\subseteq N_{-i}^{\ast}(\mathcal{C})$.
\end{lemma}

We will consider the possibility of varying the offers that a supplier makes
in a way that does not change the supply chain optimal solutions. We say that
a set $\{\Delta_{i}(\cdot): i\in N\}$ of functions from $\mathbf{T}$ to $\mathbb{R}$ is \emph{consistent} with the
cost functions $\{(C_{i}(\cdot),E_{i}(\cdot)): i\in N\}$ if adding $\Delta_{j}(\mathbf{t})$
to the cost $E_{j}(\mathbf{t})$ does not change the optimal solution or the optimal value, i.e., for every $S\subseteq N$,
\begin{equation}\label{eq:consistency}
\max\left\{\Pi_{\mathcal{C}}(\mathbf{t},\mathbf{s})+\sum_{j\in S}\Delta_{j}(\mathbf{t}):
\mathbf{t}\in\mathbf{T}, \mathbf{s}\in X(\mathbf{t}) \mbox{  for all } D \in \cal{D}, I(\mathbf{t})\subseteq S\right\}
=\Pi_{\mathcal{C}}^{\ast}(S),
\end{equation}
and the maximization is attained at $\mathbf{t}_{S}^{\ast}$. It is easy to see that this is equivalent to the conditions
of (a) $\sum_{j\in S}\Delta_{j}(\mathbf{t}_{S}^{\ast})=0$ and (b) if
$I(\mathbf{t})\subseteq S$ and $\mathbf{s}(D)\in X(\mathbf{t})$  for all $D \in \cal{D}$, then
\begin{equation}\label{ineq:consistent-variations}
\sum_{j\in S}\Delta_{j}(\mathbf{t})\leq\Pi_{\mathcal{C}}^{\ast}(S)
-\Pi_{\mathcal{C}}(\mathbf{t},\mathbf{s}).
\end{equation}
Note that when $\Delta_{i}(\mathbf{t})$ is zero for any $\mathbf{t}$, then it is trivially
consistent with the costs. Now we are ready to characterize the equilibrium that occurs when $\Pi^*_{\mathcal{C}}(S)$
is submodular as a function of $S$.

\begin{theorem}[Nash Equilibrium]\label{thm:general-equilib}
If Assumptions~\ref{assum-SCopt}--\ref{ass:submodularity} hold, then the set of bids
$\bar{\mathcal{B}}=\{(C_i(\cdot),\bar{R}_{i}(\cdot)):
i\in N^{\ast}(\mathcal{C})\}\cup \{(C_{i}(\cdot),E_{i}(\cdot)): i\notin N^{\ast}(\mathcal{C})\}$
is a Nash equilibrium, where
\[
\bar{R}_{i}(\mathbf{t})= \left\{
 \begin{array}{ll}
 E_{i}(\mathbf{t})+\Pi_{\mathcal{C}}^{\ast}(N)-\Pi_{\mathcal{C}}^{\ast}(N\backslash\{i\})-\Delta_{i}(\mathbf{t}),
      & \hbox{if $t_{i}>0$;} \\
 0, & \hbox{if $t_{i} = 0$;}
                           \end{array}
                         \right.
\]
and $\{\Delta_{i}(\cdot)\ge 0: i\in N^*(\mathcal{C})\}$ is any set of functions consistent with the costs.
In equilibrium, the buyer makes a supply chain optimal choice $(\mathbf{t}_N^*,
\mathbf{s}^*_N)$. The buyer makes profit $\Pi_{\mathcal{C}}^{\ast}(N)-\sum_{i=1}^{n}
\left(\Pi_{\mathcal{C}}^{\ast}(N)-\Pi_{\mathcal{C}}^{\ast}(N\setminus\{i\})\right)$ and supplier $i$ makes
profit $\Pi_{\mathcal{C}}^{\ast}(N)-\Pi_{\mathcal{C}}^{\ast}(N\setminus\{i\})$.
\end{theorem}

We can see immediately from Theorem~\ref{thm:general-equilib} that the following set of bids is a
Nash equilibrium (by setting all $\Delta_{i}(\mathbf{t})=0$):
\[
\bar{R}_{i}(\mathbf{t})=E_{i}(\mathbf{t})+\Pi_{\mathcal{C}}^{\ast}(N)
-\Pi_{\mathcal{C}}^{\ast}(N\backslash\{i\}), \mbox{ when } t_{i}>0,
\]
and $\bar{R}_{i}(\mathbf{t})=0$ when $t_{i}=0$.

Several points about this theorem are worth mentioning. First, the profit allocation in equilibrium is a VCG result. However,
this is not from a mechanism design that sets prices paid in a particular way, but arises as an equilibrium from our pay-as-bid
capacity game. It is straightforward that each supplier makes a nonnegative profit. Suppliers in $N(\mathcal{C})^*$ each make a
strictly positive profit while the other suppliers each make zero profits. On the buyer's side, the profit is nonnegative, which is
a direct result of the submodularity property of $\Pi_{\mathcal{C}}^*(S)$ and can also be implied by the fact that the buyer will
not purchase from any supplier should it make a negative profit.

The profit allocation is primarily driven by the level of competition between suppliers. In the extreme case where there is a
perfectly competitive market, then each supplier's
contribution is zero, since in an individual supplier's absence another supplier will step in with no overall reduction in supply
chain profits. Thus the entire supply chain profit will go to the buyer.

Another property of the equilibrium is that the buyer's profit will remain the same if any single supplier is dropped
from the set of available suppliers. We establish this in the proof of Theorem~\ref{thm:general-equilib} (see equation \eqref{eq:Bbar-values}). The intuition is that if the buyer makes a lower profit when any supplier is removed, then this supplier would have an incentive to increase its bid by a positive amount while still ensuring it is chosen by the buyer.

Finally, in equilibrium the buyer makes a choice that maximizes the supply chain profit, i.e., the supply chain is coordinated in equilibrium. It is instructive to relate our findings with those of the supply chain coordination literature \citep{Cachon03}.
In general, this literature focuses on designing sophisticated contracts (e.g., revenue sharing or buy back) to achieve the chain optimal profit in a dyadic supply chain. In contrast, we have a different supply chain structure (i.e., many to one), and
the supply chain optimality arises as a result of supplier competition (rather than by way of design).

\begin{excont}[continued]
Let us consider again the capacity game given in Example~\ref{exp:model-illustration}. If just one of the
suppliers, say $i$, is available, then it can be shown that the optimal
reservation amount for the supply chain as a whole is $t_{i}=75$ and the
supply chain profit is $\$47.5$. In this case, the amount supplied is less
than the demand in the high demand case. On the other hand, if both suppliers
are available, then it is optimal (to the supply chain) to reserve $50$ from each, so $t_{1}=t_{2}=50$.
In this case no demand is lost, but the buyer finds it
worthwhile to purchase from only one of the suppliers in the event that
demand is low (thus saving the additional \$20 cost in this case). The
supply chain profit is $\$66.67$. Thus the supply chain profits are
submodular and there is an equilibrium where each supplier offers
$P_{i}(\mathbf{s})=s_{i}$ and $R_{i}(\mathbf{t})=E_{i}(\mathbf{t})+19.17$ for
$t_{i}>0$ and $R_{i}(\mathbf{t})=0$ otherwise, where $i = 1, 2$. \ $\Box$
\end{excont}

\section{Application to capacity games with function bids}
\label{section-cap-game-functions}

We will apply our results for the general capacity game to the specific case
we described in the introduction section. The buyer faces a random demand $D$, which
follows a probability distribution $F(d)$ with density $f(d)$ over
$[\underline{d},\bar{d}]$ where $0\leq\underline{d}<\bar{d}<\infty$. For each
unit of items sold, the buyer collects a revenue $\rho$. Supplier $i$ has a
reservation cost, which is expressed as a general increasing function of the
amount reserved, with a marginal reservation cost for an amount $t$ given by
$e_{i}(t)$ (so that total reservation costs are given by $\int_{0}^{t}e_{i}(s)ds$).
In addition supplier $i$ has an execution cost which is a
general increasing function of the amount supplied, with a marginal execution
cost of $c_{i}(t)$. We have switched to lower case letters here as an
indication that these are marginal costs rather than the total costs as occur
in the general capacity game. Note that $c_{i}(t)$ and $e_{i}(t)$ could be
constants. Suppliers each maximize their own expected profits by offering a
function bid consisting of a (marginal) reservation price function $r_{i}(t)$
and a (marginal) execution price function $p_{i}(t)$ for $t\in\lbrack0,\bar
{t}]$. Without loss of generality, we assume $\bar{t}=\bar{d}$ since the buyer
will never reserve more than $\bar{d}$ units. In addition, if a supplier does
not want to offer that much, he may simply set a very high price for the
quantities beyond the desired amount so that the buyer will never choose to reserve more than the supplier wishes to offer.

Suppose the bids offered by suppliers are $\mathcal{B}=$ $\{(p_{i}(\cdot),r_{i}(\cdot)): i\in N\}$,
where the functions $p_{i}(\cdot)$ and $r_{i}(\cdot)$ are defined on $[0,\bar{d}]$ and may be constants.
The buyer makes a reservation decision in the
first stage and makes an execution decision in the second stage. Since the
reservation payment becomes sunk at the time when the buyer makes the
execution decision, with any realized demand, the buyer simply chooses the
cheapest way to meet the demand based on execution prices up to the amount
reserved from each supplier. If prices are continuous and less than $\rho$, then it is not hard to
see that the buyer will meet the demand $D$ through choosing amounts $x_{i}$
from supplier $i$ where $\sum_{i\in N} x_{i}=D$ and the prices are all equal so that we may write
$p_{i}(x_{i})=p(D)$, $i\in N$, except where $x_{j}$ is the reserved capacity
for $j$, in which case $p_{j}(x_{j})\leq p(D)$. However in the general case,
when prices can decrease, there may be more than one solution having these
properties even with continuous prices. In this case the buyer will select the
solution with the lowest total payment.

To avoid these complications and allow us to develop explicit expressions for $\Pi_\mathcal{B}$ and $\Pi_\mathcal{C}$,
we make the following simplifying assumption.

\begin{assumption}\label{assume-pi}
The execution price function $p_{i}(\cdot)$ is non-decreasing for every $i\in N$.
\end{assumption}

In order to put this problem into the general form of a capacity game, we need
to switch from marginal costs to the full costs given by their integrals. Also
we note that for the capacity game with function bids, $p_{i}$, $c_{i}$,
$r_{i}$ and $e_{i}$ only have dependence on $t_{i}$ and $s_{i}$ and not on the
full vectors $\mathbf{t}$ and $\mathbf{s}$ (so that we need to capture the
restriction that bids $\mathbf{p}$ and $\mathbf{r}$ are of this form through
the specification of the sets $A_{i}$). Thus by setting $V(D,\mathbf{s})
=\rho\min(D,\sum_{i\in N}s_{i})$, $P_{i}(\mathbf{s})=\int_{0}^{s_{i}} p_{i}(u)du$,
$C_{i}(\mathbf{s})=\int_{0}^{s_{i}}c_{i}(u)du$, $R_{i}(\mathbf{t})
=\int_{0}^{t_{i}}r_{i}(u)du$, $E_{i}(\mathbf{t})=\int_{0}^{t_{i}}e_{i}(u)du$,
we obtain the problem in the required form.

For the capacity game with function bids we can write the vector of second
stage choices $\mathbf{s}$ explicitly in terms of $\mathbf{t}$ by
recognising that the optimal choice for the buyer will use the low cost
suppliers first. As we will demonstrate below, this allows an explicit
expression for
\[
\Pi_{\mathcal{B}}(\mathbf{t}):=\max_{\mathbf{s}\leq\mathbf{t}}\Pi_{\mathcal{B}}(\mathbf{t},\mathbf{s})
 =\max_{\mathbf{s}\leq\mathbf{t}}\mathbb{E}\left[\rho\min(D,\sum_{i\in N}\mathbf{s}(D)_{i})
 -\sum_{i\in N}(P_{i}(\mathbf{s}(D)_{i})+R_{i}(t_{i}))\right],
\]
where $\mathcal{B}=\{(P_{i}(\cdot),R_{i}(\cdot)),i\in N\}$ and we write $\mathbf{s}(D)_i$ for the $i$th
component of $\mathbf{s}(D)$. Similarly we define $\Pi_{\mathcal{C}}(\mathbf{t}):=\max_{\mathbf{s}\leq
\mathbf{t}}\Pi_{\mathcal{C}}(\mathbf{t},\mathbf{s})$. Let
\begin{equation}
\gamma_{j}(p)=\sup\{0\leq y\leq t_{j}:p_{j}(y)\leq p\}. \label{eqn-lambda}%
\end{equation}
Then $\gamma_{j}(p_{i}(x))$ is the amount purchased from supplier $j\neq i$ if
an amount $x$ is purchased from supplier $i$. When $p_{j}(t)$ is strictly
increasing and continuous, then $\gamma_{j}(p)=\min\{p_{j}^{-1}(p),t_{j}\}$.
In the special case where marginal execution costs are constant (i.e.,
$p_{k}(t)=p_{k}$ for all $k\in N$ where $p_{k}$ is a constant), we have
$\gamma_{j}(p)=0$ if $p_{j}>p$, and $\gamma_{j}(p)=t_{j}$ if $p_{j}\leq p$.

Let $\mathbf{t}_{-i}=(t_{1},\cdots,t_{i-1},t_{i+1},\cdots,t_{n})$ denote the
vector of reservation choices from suppliers other than $i$. We now write the
cumulative amount of capacity with execution prices less than $p_{i}(x)$ as
follows,
\begin{equation}\label{eq-h}
h_{i}(x,\mathbf{t}_{-i})=x+\sum_{j\neq i}\gamma_{j}(p_{i}(x)),\quad
x\in\lbrack0,t_{i}],
\end{equation}
where $\gamma_{j}$ is defined in (\ref{eqn-lambda}), the term $x$ indicates
the dispatched amount from supplier $i$, and the second term represents the
dispatched amount from suppliers other than $i$. If the buyer chooses
optimally then this is the total amount that the buyer purchases conditional on her purchasing amount
$x$ from supplier $i$. Hence with optimal buyer choices, the amount purchased from
supplier $i$ is greater than $x$ if and only if the demand is greater than
$h_{i}(x,\mathbf{t}_{-i})$.

Thus, with the reservation choice $\mathbf{t}$, the buyer's optimal expected
profit is
\begin{equation}\label{eqn-buyer-profit-n}
\Pi_{\mathcal{B}}(\mathbf{t})=\sum_{i=1}^{n}\int_{0}^{t_{i}}\left\{
[\rho-p_{i}(x)]\bar{F}(h_{i}(x,\mathbf{t}_{-i}))-r_{i}(x)\right\}dx,
\end{equation}
where $\mathcal{B}=$ $\{(p_{i}(t),r_{i}(t)):i\in N\}$, $h_{i}(\cdot,\cdot)$ is
given in (\ref{eq-h}) and $\bar{F}(z)=\text{Pr}[D\geq z]$. We know that
$\bar{F}(h_{i}(x,\mathbf{t}_{-i}))$ represents the probability that the buyer
will use the $x$th unit of the reserved capacity from supplier $i$, and thus
each term within the summation gives the buyer's profit of reserving $t_{i}$
from supplier $i$. Similarly for the supply chain as a whole we define
\begin{equation}\label{eq-hbar}
\bar{h}_{i}(x,\mathbf{t}_{-i})=x+\sum_{j\neq i}\sup\{0\leq y\leq t_{j}:
c_{j}(y)\leq c_{i}(x)\},\quad x\in\lbrack0,t_{i}],
\end{equation}
and we have
\begin{equation}\label{eqn-supply-chain-profit-n}
\Pi_{\mathcal{C}}(\mathbf{t})=\sum_{i=1}^{n}\int_{0}^{t_{i}}\left\{
(\rho-c_{i}(x))\bar{F}(\bar{h}_{i}(x,\mathbf{t}_{-i}))-e_{i}(x)\right\}dx.
\end{equation}

Now we apply Theorem~\ref{thm:best-resp} to get the following corollary in
this case.

\begin{corollary}\label{Corr-br-n}
Given bids $\{(p_{j}(\cdot),r_{j}(\cdot)):j\in N,j\neq i\}$ of the other suppliers, if supplier $i$ has preferred status,
then it is optimal for supplier $i$ to set $p_{i}(t)=c_{i}(t)$ for $t\in\lbrack0,\bar{d}]$
and choose the function $r_{i}(\cdot)$ such that
\[
\int_{0}^{\hat{\mathbf{t}}_{i}}r_{i}(t)dt=\int_{0}^{\hat{\mathbf{t}}_{i}}e_{i}(t)dt
+\Pi_{\mathcal{B}_{i}}^{\ast}(N)-\Pi_{\mathcal{B}_{i}}^{\ast}(N\backslash\{i\}),
\]
where $\mathcal{B}_{i}=\{(p_{j}(\cdot),r_{j}(\cdot)),j\in N,j\neq i\}\cup \{(c_{i}(\cdot),e_{i}(\cdot))\}$ and
$\hat{\mathbf{t}}\in\arg\max\Pi_{\mathcal{B}_{i}}(\mathbf{t})$ provided $r_{i}(\cdot)$ is chosen in such a way that the optimal
buyer choice is $\hat{\mathbf{t}}$. An optimal choice for supplier $i$ is to choose $r_{i}(t)=e_{i}(t)$ for all $t\in [0,\bar{d}]$
and in addition charge a fixed amount of $\Pi_{\mathcal{B}_{i}}^{\ast}(N)-\Pi_{\mathcal{B}_{i}}^{\ast}(N\backslash \{i\})$ for any
non-zero amount reserved. In the case where supplier $i$ does not have preferred status, then it is possible for
supplier $i$ to achieve within $\epsilon$ of the best profit by choosing the same bid of $p_i(\cdot)$ and a bid of $r_i(\cdot)$
that is reduced by $\epsilon$.
\end{corollary}

This result shows that for the capacity game with function bids there is an
optimal solution with $p_{i}(t)=c_{i}(t)$ for the supplier $i$'s best response
problem, but there will be multiple possible solutions for $r_{i}(t)$ such
that the buyer's optimal choice is $\hat{\mathbf{t}}$ and the supplier makes a
profit of $\Pi_{\mathcal{B}_{i}}^{\ast}(N)-\Pi_{\mathcal{B}_{i}}^{\ast
}(N\backslash\{i\})$ (which will imply that the buyer makes the same profit as
$\Pi_{\mathcal{B}_{i}}^{\ast}(N\backslash\{i\})$). In all of these solutions,
the optimal buyer choice is $\hat{\mathbf{t}}$, which maximizes the buyer's
profit when supplier $i$ charges only his costs.

For the capacity game with function bids we also have the following immediate corollary
from Theorem~\ref{thm:general-equilib}.

\begin{corollary}[Nash equilibrium with a lump sum payment]\label{Cor-lumpsum-n}
Suppose Assumptions~\ref{assum-SCopt}--\ref{ass:submodularity} hold. Then there exists a Nash equilibrium
$\{(p_{i}^{\ast}(\cdot),r_{i}^{\ast}(\cdot),K_{i}^{\ast}): i\in N\}$, where each
supplier $i$ sets prices to be costs, i.e., $p_{i}^{\ast}(\cdot)=c_{i}(\cdot)$,
$r_{i}^{\ast}(\cdot)=e_{i}(\cdot)$, and charges a lump-sum reservation payment
$\Pi_{\mathcal{C}}^{\ast}(N)-\Pi_{\mathcal{C}}^{\ast}(N\backslash\{i\})$. The
buyer's reservation choice is $\mathbf{t}_{N}^{\ast}$, the buyer makes a profit of
$\Pi_{\mathcal{C}}^{\ast}(N)-\sum_{i=1}^{n}\Pi_{\mathcal{C}}^{\ast}(N)-\Pi_{\mathcal{C}}^{\ast}(N\backslash\{i\})$,
and supplier $i$ makes a profit of $\Pi_{\mathcal{C}}^{\ast}(N)-\Pi_{\mathcal{C}}^{\ast}(N\backslash\{i\})$. \ $\Box$
\end{corollary}

One property of the equilibrium involving a lump sum payment is that it cannot
be represented simply by defining marginal prices for capacity: to do so would
require an infinite price for the first $\varepsilon$ of capacity. However, we
can use the functions $\Delta_{i}(\mathbf{t})$ occurring in Theorem~\ref{thm:general-equilib}
to construct an equilibrium solution with finite
marginal costs. This is done by adjusting the capacity offer from the lump sum
bid by smoothing out the beginning of the offer. We will make use of power
functions to do this.

\begin{proposition}[Nash equilibrium with power functions]\label{prop:equilibrium-power}
Suppose Assumptions~\ref{assum-SCopt}--\ref{ass:submodularity} hold. Further we assume that $c_{i}(\cdot)$ and
$e_{i}(\cdot)$ are non-decreasing. Then the following strategies form a Nash
equilibrium: For each $i\in N$, $p_{i}^{\ast}(\cdot)=c_{i}(\cdot)$ and
\[
r_{i}^{\ast}(t)=\left\{
\begin{array}[c]{ll}
e_{i}(t)+\delta_{i}(t;\beta_{i}),  & \text{ if } 0\leq t<\theta_{i},\\
e_{i}(t), & \text{ if } \theta_{i}\leq t\leq\bar{d},
\end{array}
\right.
\]
where
\[
\delta_{i}(t;\beta_{i})=(\Pi_{\mathcal{C}}^{\ast}(N)-\Pi_{\mathcal{C}}^{\ast}
(N\setminus\{i\}))\frac{(\beta_{i}+1)}{\theta_{i}}\left(\frac{\theta_{i}-t}{\theta_{i}}\right)^{\beta_{i}},
\]
$\theta_{i}=\min((\mathbf{t}_{S}^{\ast})_{i}: S\subseteq N, S \ni i)$, and $\beta_{i}>1$ is a constant large enough. In
equilibrium, the buyer's reservation choice is $\mathbf{t}_{N}^{\ast}$. The buyer makes a profit of
$\Pi_{\mathcal{C}}^{\ast}(N)-\sum_{i=1}^{n}\left(\Pi_{\mathcal{C}}^{\ast}(N)-\Pi_{\mathcal{C}}^{\ast}(N\setminus\{i\})\right)$
and supplier $i$ makes a profit of $\Pi_{\mathcal{C}}^{\ast}(N)-\Pi_{\mathcal{C}}^{\ast}(N\setminus\{i\})$.
\end{proposition}

This construction involves adding the margin $\delta_{i}(t;\beta_{i})$
to the reservation costs, which is a decreasing function of
$t\in (0,\theta_{i})$. Choosing a high value for $\beta_{i}$ will imply that
$\delta_{i}(t;\beta_{i})$ decreases more steeply at the beginning and becomes
flatter at the end of this interval.

From Proposition~\ref{prop:equilibrium-power}, we see that there is a continuum of Nash equilibria with power functions, each
associated with a possible set of $\beta_i$ values. We can show that when $\beta_i$ approaches positive infinity, the power function
bid reduces to the lump sum bid. Thus, Corollary~\ref{Cor-lumpsum-n} can be considered as a limiting example of
Proposition~\ref{prop:equilibrium-power}. Moreover, one may easily construct an equilibrium where some suppliers use lump sum bids
(with infinitely large $\beta$ values) while others use power function bids. Despite the multiplicity of equilibrium bidding
strategies, all these equilibria lead to the same profit allocation and thus the equilibrium outcome characterized in
Proposition~\ref{prop:equilibrium-power} is essentially unique.

\subsubsection*{Comparison with MS}

As discussed earlier, existing models have focused on the case where the marginal costs are constant. As we will show in
Section~\ref{sec:submodularity}, the supply chain optimal profit is submodular for this special case, and hence the equilibria
characterized in Corollary~\ref{Cor-lumpsum-n} and Proposition~\ref{prop:equilibrium-power} apply. It is interesting to draw a
parallel between our model and the existing ones for this case. To this end, we draw on an example from MS to demonstrate the
difference in equilibrium outcomes.

\begin{example}[Example 1 in MS]\label{example-sc}
The buyer's demand is uniformly distributed over $[0,1]$, so $F(t) = t$ for $t\in[0,1]$. There are two
suppliers and their marginal costs are $(c_{1}, e_{1}) = (0,60)$ and $(c_{2}, e_{2} ) = (75,5)$. The retail price is
$\rho= 100$. We now examine the supply chain optimal problems.

If the buyer reserves from two suppliers, the supply chain problem is:
\[
\max_{t_{1},\,t_{2} \in[0, 1]} \left\{  \int_{0}^{t_{1}} \left(  (\rho-c_{1}) \bar{F}(x) - e_{1} \right)
dx +\int_{0}^{t_{2}} \left(  (\rho-c_{2}) \bar{F}(x+ t_{1}) - e_{2} \right)
dx \right\}.
\]
The optimal solution is $\mathbf{t}^*_N = (4/15, 8/15)$. The supply chain optimal profit is $\Pi^*_\mathcal{C}(N)= 32/3$.

If the buyer chooses supplier $1$ only, the supply chain problem is:
\[
\max_{t_{1} \in[0, 1] } \left\{  \int_{0}^{t_{1}} \left(  (\rho-c_{1}) \bar{F}(x) - e_{1} \right) dx\right\}.
\]
The optimal solution is $\mathbf{t}^*_{\{1\}} = (2/5)$ and the supply chain optimal
profit is $\Pi^*_\mathcal{C}(\{1\}) = 8$.

If the buyer chooses supplier $2$ only, the supply chain problem is:
\[
\max_{t_{2} \in[0, 1]} \left\{  \int_{0}^{t_{2}} \left(  (\rho-c_{2}) \bar{F}(x) - e_{2} \right)dx \right\}.
\]
The optimal solution is $\mathbf{t}^*_{\{2\}} = (4/5)$ and the supply chain optimal
profit is $\Pi^*_\mathcal{C}(\{2\}) = 8$.

If suppliers are restricted to each offering a pair of reservation price and execution price, in equilibrium these two
suppliers bid infinitesimally close to each other. MS show that the following
is a continuum of $\epsilon$-equilibria, which are parameterized with
$p\in[50, 75]$:
\[
(p_{1}^{*}, r_{1}^{*}) = (p_{2}^{*}, r_{2}^{*}) = \left(  p, 60 - \frac{55}{75}p \right).
\]
In equilibria, the buyer's reservation choice is
\[
\mathbf{t}^* = \left(\frac{4}{15},\frac{40}{3(100-p)}\right).
\]
The profit split amongst players is given by
\begin{equation}\label{eqn:example-profit_split1}
  \Pi_{B}^{*} = \frac{8(150-p)^{2}}{225(100-p)} , \quad\pi_{1}^{*} =
\frac{8p}{225}, \quad\pi_{2}^{*} = \frac{800 (75-p)}{9 (100-p)^{2}},
\end{equation}
and the supply chain profit is given by
\[
\Pi_\mathcal{C}^{*} = \frac{32 (225- 2p)}{9 (100-p)} + \frac{800 (75 - p)}{9 (100-p)^{2}}.
\]

Note that all the above equilibria are inefficient (i.e., not supply chain
optimal) except the one with $p=75$. At this efficient equilibrium, each
supplier offers a bid $(75,5) $ which is identical to the supplier $2$'s cost.
The supplier $2$'s profit is $0$, the supplier $1$'s profit is $8/3$, and the
buyer's profit equals $8$.

We now demonstrate that the above strategies do not form an equilibrium if we
allow suppliers to offer function bids. Suppose supplier $1$ chooses the
proposed bid $(p_{1}^{*},r_{1}^{*}) = (p, 60 - (55/75)p)$, and we now examine
supplier $2$'s best response in choosing a function bid.

First, if supplier $1$ is the sole supplier, the buyer's reserved amount will
be the sum of the two components of $\mathbf{t}^{*}$, and the buyer's profit is equal to
$\Pi_{B}^{*}$. Therefore, we have
\[
\mathbf{t}_{\{1\}}^{*} = \frac{600 - 4p}{15(100-p)} \quad\text{ and } \quad\Pi_{B}^{*}(\{1\}) =
\frac{8(150-p)^{2}}{225(100-p)}.
\]

Second, we show that the following strategy for supplier $2$ improves his
profit: setting prices to be costs and charging a lump-sum payment of $\frac{32(75 - p)}{9(100-p)}$.
Given this offer (and the supplier $1$'s offer $(p_{1}^{*}, r_{1}^{*})$), we
can show that the solution for the buyer's problem is
\[
\tilde{\mathbf{t}} = \left(\frac{4}{15}, \frac{8}{15}\right),
\]
and the buyer's profit from choosing $\tilde{\mathbf{t}}$ is
equal to $\Pi_{B}^{*}(\{1\})$. Also if the buyer purchases from only supplier $2$,
the buyer makes a profit of $\Pi_{B}^{*}(\{1\})$ as well. According to the
tie-breaking rule, the buyer will select $\tilde{\mathbf{t}}$.
Then supplier 2's profit becomes
\[
\tilde{\pi}_{2} = \frac{32(75 - p)}{9(100-p)} ,
\]
which is strictly greater than $\pi_{2}^{*}$ for any $p < 75$. This shows
that the equilibria in MS do not hold if we allow suppliers to offer a
function bid.

Under our model setup, the problem is well behaved and there is an equilibrium with lump-sum payments (and there is also an
equilibrium with power functions).
The following bids form a Nash equilibrium  (where $K_1$ and $K_2$ are lump sum payments):
\[
\left(p_{1}^{*}, r_{1}^{*}, K_{1}\right) =\left( 0, 60, \frac{8}{3}\right) \quad\text{ and }
\quad \left(p_{2}^{*}, r_{2}^{*}, K_{2} \right) = \left(75, 5, \frac{8}{3}\right).
\]
At this equilibria, the buyer's optimal reservation choice is $\mathbf{t}^*= ( 4/15, 8 / 15 )$.
The profit split amongst players is given by
\begin{equation}\label{eqn:example-profit_split2}
  \Pi_{B}^{*} = \frac{16}{3} \quad\text{ and } \quad\pi_{1}^{*} = \pi_{2}^{*} =
\frac{8}{3}.
\end{equation}
In this equilibria, each supplier's profit equals his contribution to the
supply chain system and the buyer takes the remaining profit. Moreover, the reservation choice
by the buyer is supply chain optimal (in distinction to the case with constant prices).

\medskip

The key message of Example~\ref{example-sc} is that imposing the restriction that each
supplier submits a pair of reservation price and an execution price rather
than a function bid leads to a higher buyer profit. On the other hand, the suppliers are better off if they submit
function bids. This can be easily seen by comparing the profit splits at \eqref{eqn:example-profit_split1} and
\eqref{eqn:example-profit_split2}. \ $\Box$
\end{example}

\subsubsection*{Comparison with ACS}

In our setting, each supplier offers a function bid with a marginal
reservation price function and a marginal execution price function, and the
buyer has the freedom to reserve whatever amount she likes. This setting
resembles the two extensions to the setting with blocks of capacity discussed in \cite{AndersonChenShao2017}.
The two extensions they discussed were:

(a) \textit{Partial Reservation}, in which the buyer is not restricted to
reserving a whole block. Each supplier owns a single block and the blocks can
be of different sizes. Every supplier chooses an execution price and a
reservation price that apply to all elements of his block.

(b) \textit{Multiple Blocks with Common Owner}, in which each supplier owns
multiple unit-blocks and can choose different prices for different unit
blocks. The buyer can freely choose among the offered blocks.

In both extensions, the authors demonstrate that one of their key results in the
baseline model, $p_{i} = c_{i}$ for supplier $i$'s best response, does not
hold. In contrast, we show that the result $p_{i} = c_{i}$ holds in our
setting as shown in Corollary~\ref{Corr-br-n}.

The key differences are as follows. In the case (a) of partial reservation, the buyer can freely reserve any amount
as in our setting. However, suppliers are restricted to offering just two scalar
prices rather than general function bids.
In the case (b) of multiple blocks with common owner, each supplier can choose different prices for
different blocks of capacity as in our setting. However, the buyer can freely
choose any of the blocks, whereas in our function bidding setting, the buyer does not have as much flexibility,
because the buyer will need to accept the prices for the
earlier blocks in order to enjoy the prices for the next blocks.
These differences in results highlight the importance of each player's choice flexibility on the equilibrium outcomes.

\section{Discussion of submodularity assumption}
\label{sec:submodularity}

In this section we examine two commonly studied settings in supply chains: (a) there are only two suppliers
competing for the buyer's procurement business, and (b) there are more than two suppliers and the marginal execution costs
are constants. In each case we are interested in establishing conditions that will ensure submodularity of
${\Pi}_{\mathcal{C}}^{\ast}(S)$.

\subsection{Case with two suppliers}

When there are only two suppliers (denoted by $1$ and $2$) we can find
conditions that are enough to ensure that submodularity occurs, and it turns
out that in the special case of the capacity game with function bids submodularity will hold
without any further restrictions.

\begin{proposition}[Subadditivity]\label{prop-subadditive}
If $V(D,\mathbf{s})$ is subadditive in $\mathbf{s}$ and the cost functions are defined separately on each
component, so $C_{i}(\mathbf{s})$ is determined by the $i$-th component of $\mathbf{s}$, and $E_{i}(\mathbf{t})$
is determined by the $i$-th component of $\mathbf{t}$, then ${\Pi}_{\mathcal{C}}^{\ast}(S)$ is submodular in $S\subseteq N$.
In the case of the capacity game with function bids, these conditions will hold.
\end{proposition}

\subsection{Case with constant marginal execution costs}

In the case of more than two suppliers, the submodularity of supply chain profit
may not hold in general, as we show in Example~\ref{exp:non-submodularity} below.
The following theorem identifies some sufficient conditions for the submodularity property
and these conditions are satisfied for a large class of the capacity games with
function bids. The notion of laminar convexity stated in the theorem is defined in the Appendix.

\begin{theorem}[Submodularity]\label{thm:submodular}
Let $\mathbf{T} =[0,\bar{d}]^n$ and $X(\mathbf{t})=\{\mathbf{s}\in \mathbb{R}_+^n: \mathbf{s}
\le \mathbf{t}\}$ for $\mathbf{t}\in \mathbf{T}$. If $\max_{\mathbf{s}\in X(\mathbf{t})}
\left(V(D,\mathbf{s})-\sum_{i=1}^{n}C_{i}(\mathbf{s})\right)$ is laminar concave in $\mathbf{t}$
and each cost function $E_{i}(\mathbf{t})$ ($i\in N$) is a convex function of (only) the $i$-th component of $\mathbf{t}$,
then ${\Pi}_{\mathcal{C}}^{\ast}(S)$ is submodular in $S\subseteq N$. In the case
of the capacity game with function bids, these conditions are satisfied if for each supplier the marginal
execution cost is constant and the marginal reservation cost is non-decreasing.
\end{theorem}

The complexity of the analysis arises from the fact that the supply chain profit function cannot be easily decomposed into
components related to individual suppliers. The profit made from supplier $i$ is related to the capacity reservations from
all the suppliers with lower execution prices.

Theorem~\ref{thm:submodular} implies that, in settings considered by the existing studies where the marginal costs are
constant, the supply chain optimal profit is submodular. We use the following example to illustrate this point.

\begin{example}
Suppose the buyer demand $D$ follows a uniform distribution over $[0,1]$.
There are three suppliers whose costs are: $c_{1} = 1, e_{1} = 3$; $c_{2} =
2.5, e_{2} = 2$; and $c_{3} = 5, e_{3} = 1$. The retail price is $\rho= 10$.
We know from Theorem~\ref{thm:submodular} that the problem is submodular,
and we can carry out the detailed calculations to find the supply chain optimal
solutions for different sets of available suppliers. One way to do these calculations is to use the
screening curve approach that is common in calculation of optimal generation mix in electricity markets
\citep[see][]{Green05}. We summarize the optimal reservation choices and profits in Table~\ref{tab-sc-results}.

\renewcommand*{\arraystretch}{1.2}
\begin{table}[th]
\centering
\begin{tabular}[c]{c|c|c|c|c}\hline
Available Suppliers $S$ & $(\mathbf{t}^*_S)_1$ & $(\mathbf{t}^*_S)_{2}$ & $(\mathbf{t}^*_S)_{3}$
 & $\Pi_{\mathcal{C}}^*(S)$\\\hline
$\{1,2,3\}$ & $1/3$ & $4/15$ & $1/5$ & $2.1333$\\
$\{1,2\}$ & $1/3$ & $2/5$ & $0$ & $2.1$\\
$\{1,3\}$ & $1/2$ & $0$ & $3/10$ & $2.1$\\
$\{2,3\}$ & $0$ & $3/5$ & $1/5$ & $2.05$\\
$\{1\}$ & $2/3$ & $0$ & $0$ & $2$\\
$\{2\}$ & $0$ & $11/15$ & $0$ & $2.0167$\\
$\{3\}$ & $0$ & $0$ & $4/5$ & $1.6$\\ \hline
\end{tabular}
\caption{The supply chain optimal reservation choices and profits}\label{tab-sc-results}
\end{table}
Using the results in the table, we can easily check the submodularity of $\Pi^*_{\mathcal{C}}(S)$.
We can now construct an equilibrium set of offers where the suppliers offer at cost and in addition
require a lump sum reservation payment of $\Pi^*_\mathcal{C}(N) - \Pi^*_\mathcal{C}(N\backslash\{i\})$,
which then becomes the supplier profit. Here these amounts are $0.08333$ for supplier 1, $0.0333$ for
supplier 2, and $0.0333$ for supplier 3. In this equilibrium the buyer receives the remainder of the
total supply chain profit: $2.1333-0.15=1.9833$. \ $\Box$
\end{example}

The submodularity property may not hold when suppliers have decreasing
marginal costs as we demonstrate with the following example.

\begin{example}\label{exp:non-submodularity}
Suppose the demand is fixed with $D = 10$ and
the retail price is $\rho= 20$. There are three suppliers with $N = \{1,2,3\}$.
Supplier $1$ and supplier $2$ have the same costs
with $c_{1}(t) = c_{2}(t) = e_{1}(t) = e_{2}(t) = 0 $, for $t \in[0, 5]$ (and
an infinite cost for any larger amount). Supplier $3$'s costs are $c_{3}(t) = 0$ and $e_{3}(t) = 10 - t$ for $t
\in[0,10] $. So both supplier $1$ and supplier $2$ have the capacity of $5$
and supplier $3$'s capacity is $10$.

We now look at the supply chain optimal problems. If all the three suppliers
are available, the buyer will choose $5$ units from each of supplier $1$ and
supplier $2$. The supply chain optimal profit is $\Pi(\{1,2,3\}) = 200$. If
only suppliers $3$ and $1$ (or $2$) are available, the buyer will choose $5$
units from each of $3$ and $1$ (or $2$). The supply chain optimal profit is
$\Pi(\{1,3\}) = \Pi(\{2,3\}) = 162.5$. If supplier $3$ is the sole supplier,
the buyer will choose $10$ units from supplier $3$ and the supply chain
optimal profit is $\Pi( \{3\})= 150$. Therefore, we have $\Pi(\{1,2,3\})+ \Pi(
\{3\}) = 350 > 325 = \Pi(\{1,3\}) + \Pi( \{2,3\}), $ which contradicts the
submodularity property.


We can also show that the proposed equilibrium structure will not apply in this case.
If each of suppliers 1 and 2 asks for a lump-sum payment of $200 - 162.5 = 37.5$, and
the buyer makes the supply chain optimal choice of selecting these two suppliers, then
the buyer profit is $200 - 75=125$. However, this is less than the profit available to
the buyer from selecting supplier 3 alone, which gives the buyer $150$ as profit. So this
is not an equilibrium. We can check that an equilibrium exists where both suppliers 1 and 2
ask for a lump-sum payment of $25$, and the buyer chooses both of these offers. \ $\Box$
\end{example}

\section{Discussion and conclusions}
\label{section-conclusion}

In this paper, we have developed a general framework to study a broad class of capacity games.
This framework allows us to examine supplier competition in an option market
where suppliers' costs may be nonlinear. In this setting, suppliers each submit a function bid
consisting of a reservation price function and an execution price function.
The buyer decides how much capacity to reserve from each supplier before
knowing the actual demand.

When the competitors' bids are observed, we have shown that an optimal strategy for each
supplier is to set the execution price to be the execution cost and add a
margin on the reservation cost. This implies that suppliers make profits only
from the buyer's reservation payments.
This result does not hold in the case of bids of constant marginal costs (considered by MS).

We have also shown that, under the assumption that the supply chain optimal profit is submodular in the set of
available suppliers, there is a class of equilibria in which the buyer's reservation choice is
first best, each supplier's profit equals his marginal contribution to the
supply chain and the buyer takes the remaining profit. The implication is that
by allowing suppliers to compete using function bids, the supply chain is
coordinated.

To demonstrate the generality of the submodularity assumption, we have shown that the supply chain
optimal profit is indeed submodular when each supplier's marginal execution cost is
constant, a setting that has been studied extensively in the existing literature, or when there are only two suppliers.

From an electricity market perspective, the profits for the buyer correspond to the
overall consumer welfare. In this context the capacity mechanism we have described combines the capacity auction and the market
for energy into a single pay-as-bid auction, rather than having separate uniform price auctions. This achieves an efficient
outcome at equilibrium, even in the case where generators can exercise market power.
In the special case of a competitive environment when no single generator has a significant impact
on overall system welfare when removed, we have found that each generator bids at cost and we retain the property of an efficient
set of capacity and generation choices. The result we obtain on best response is
interesting from the perspective of electricity capacity mechanisms since it demonstrates
that when generators are required to specify their energy bids in advance with pay-as-bid
in the spot, then there is no incentive to use market power in the energy component of the
bids, with profits being made entirely from the capacity payments. This result applies
without an assumption of monotone prices and allows quite general dispatch mechanisms.
This has relevance to the issues of uplift payments that occur in US wholesale markets because of no-load and start-up costs
\citep[see, for example,][]{Hogan14}.


This paper can be extended in several directions. First we assume, as in other
supplier competition models, that supplier costs are known to the other
suppliers. This assumption fits some settings better than others. For example,
in the electricity market, generators tend to know each other's generation
technologies; thus it is prudent to assume complete cost information. However,
in other settings a model that considers cost
uncertainties may be more appropriate; see supply function equilibrium models
with private information \citep{vives2011}. In addition, our model focuses on
the contract market only, and incorporation of a spot market would be another
interesting direction.
Our equilibrium analysis builds on the submodularity condition. Even though in
Example~\ref{exp:non-submodularity} we demonstrate that, when the submodularity property fails, the VCG strategies are not a Nash equilibrium, it would be interesting to investigate further what the equilibrium looks like when the supply chain optimal profit is not submodular.
Finally, we should note that our assumption is that the random demand is exogenous.
An important extension that we do not consider is the case where the buyer can influence demand through
setting a price.


\bibliographystyle{apa}
\bibliography{Bibliography}

\begin{thebibliography}{}

\bibitem[\protect\astroncite{Anderson et~al.}{2017}]{AndersonChenShao2017}
Anderson, E.~J., Chen, B., and Shao, L. (2017).
\newblock Supplier competition with option contracts for discrete blocks of
  capacity.
\newblock {\em Operations Research}, 65(4):952--967.

\bibitem[\protect\astroncite{Anderson and Hu}{2008}]{anderson2008}
Anderson, E.~J. and Hu, X. (2008).
\newblock Finding supply function equilibria with asymmetric firms.
\newblock {\em Operations Research}, 56(3):697--711.

\bibitem[\protect\astroncite{Anderson and Philpott}{2002}]{anderson2002}
Anderson, E.~J. and Philpott, A.~B. (2002).
\newblock Optimal offer construction in electricity markets.
\newblock {\em Mathematics of Operations Research}, 27(1):82--100.

\bibitem[\protect\astroncite{Anton and Yao}{1989}]{Anton1989}
Anton, J.~J. and Yao, D.~A. (1989).
\newblock Split awards, procurement, and innovation.
\newblock {\em The RAND Journal of Economics}, 20(4):538--552.

\bibitem[\protect\astroncite{Anton and Yao}{1992}]{Anton1992}
Anton, J.~J. and Yao, D.~A. (1992).
\newblock Coordination in split award auctions.
\newblock {\em The Quarterly Journal of Economics}, 107(2):681--707.

\bibitem[\protect\astroncite{Ausubel and Milgrom}{2006}]{AusMil06}
Ausubel, L. and Milgrom, P. (2006).
\newblock The lovely but lonely {V}ickrey auction.
\newblock {\em Combinatorial Auctions}, 17:22--26.

\bibitem[\protect\astroncite{Barnes-Schuster et~al.}{2002}]{Yehuda2002}
Barnes-Schuster, D., Yehuda, B., and Anupindi, R. (2002).
\newblock Coordination and flexibility in supply contracts with options.
\newblock {\em Manufacturing \& Service Operations Management}, 4(3):171--207.

\bibitem[\protect\astroncite{Bernheim and Whinston}{1986}]{bernheim1986}
Bernheim, B.~D. and Whinston, M.~D. (1986).
\newblock Menu auctions, resource allocation, and economic influence.
\newblock {\em The Quarterly Journal of Economics}, 101(1):1--32.

\bibitem[\protect\astroncite{Bresky}{2013}]{Bresky2013205}
Bresky, M. (2013).
\newblock Revenue and efficiency in multi-unit uniform-price auctions.
\newblock {\em Games and Economic Behavior}, 82:205--217.

\bibitem[\protect\astroncite{Burnetas and Ritchken}{2005}]{Burnetas2005}
Burnetas, A. and Ritchken, P. (2005).
\newblock Option pricing with downward-sloping demand curves: The case of
  supply chain options.
\newblock {\em Management Science}, 51(4):566--580.

\bibitem[\protect\astroncite{Bushnell and Oren}{1994}]{BusOre94}
Bushnell, J. and Oren, S. (1994).
\newblock Bidder cost revelation in electric power auctions.
\newblock {\em Journal of regulatory economics}, 6(1):5--26.

\bibitem[\protect\astroncite{Cachon}{2003}]{Cachon03}
Cachon, G.~P. (2003).
\newblock Supply chain coordination with contracts.
\newblock In Graves, S.~C. and de~Kok, A.~G., editors, {\em Handbooks in
  operations research and management science: Supply chain management},
  chapter~7, pages 229--339. North Holland, Amsterdam, The Netherlands.

\bibitem[\protect\astroncite{Chao and Wilson}{2002}]{ChaWil02}
Chao, H. and Wilson, R. (2002).
\newblock Multi-dimensional procurement auctions for power reserves: Robust
  incentive-compatible scoring and settlement rules.
\newblock {\em Journal of Regulatory Economics}, 22(2):161--183.

\bibitem[\protect\astroncite{Fu et~al.}{2010}]{Fu2010}
Fu, Q., Lee, C.-Y., and Teo, C.-P. (2010).
\newblock Procurement management using option contracts: Random spot price and
  the portfolio effect.
\newblock {\em IIE Transactions}, 42(11):793--811.

\bibitem[\protect\astroncite{Green}{2005}]{Green05}
Green, R. (2005).
\newblock Electricity and markets.
\newblock {\em Oxford Review of Economic Policy}, 21(1):67--87.

\bibitem[\protect\astroncite{Ha et~al.}{2011}]{HaTonZha11}
Ha, A.~Y., Tong, S., and Zhang, H. (2011).
\newblock Sharing demand information in competing supply chains with production
  diseconomies.
\newblock {\em Management Science}, 57(3):566--581.

\bibitem[\protect\astroncite{Haldi and Whitcomb}{1967}]{Haldi&Whitcomb67}
Haldi, J. and Whitcomb, D. (1967).
\newblock Economies of scale in industrial plants: Part 1.
\newblock {\em Journal of Political Economy}, 75(4):373--385.

\bibitem[\protect\astroncite{Hobbs et~al.}{2000}]{Hobbs_et_al2000}
Hobbs, B.~F., Rothkopf, M.~H., Hyde, L.~C., and O'{N}eill, R.~P. (2000).
\newblock Evaluation of a truthful revelation auction in the context of energy
  markets with nonconcave benefits.
\newblock {\em Journal of Regulatory Economics}, 18(1):5--32.

\bibitem[\protect\astroncite{Hogan}{2014}]{Hogan14}
Hogan, W. (2014).
\newblock Electricity market design and efficient pricing: Applications for
  {N}ew {E}ngland and beyond.
\newblock {\em The Electricity Journal}, 27(7):23--49.

\bibitem[\protect\astroncite{Johari and Tsitsiklis}{2011}]{Johari2011}
Johari, R. and Tsitsiklis, J.~N. (2011).
\newblock Parameterized supply function bidding: Equilibrium and efficiency.
\newblock {\em Operations Research}, 59(5):1079--1089.

\bibitem[\protect\astroncite{Klagnanam and Parkes}{2004}]{Kalagnanam2004}
Klagnanam, J. and Parkes, D.~C. (2004).
\newblock Auctions, bidding and exchange design.
\newblock In Simchi-Levi, D., Wu, S.~D., and Shen, Z.~M., editors, {\em
  Handbook of Quantitative Supply Chain Analysis: Modeling in the E-Business
  Era}, chapter~5, pages 143--212. Springer, Boston, MA., USA.

\bibitem[\protect\astroncite{Kleindorfer and Wu}{2003}]{Kleindorfer2003}
Kleindorfer, P.~R. and Wu, D.~J. (2003).
\newblock Integrating long- and short-term contracting via business-to-business
  exchanges for capital-intensive industries.
\newblock {\em Management Science}, 49(11):1597--1615.

\bibitem[\protect\astroncite{Klemperer and Meyer}{1989}]{klemperer1989}
Klemperer, P.~D. and Meyer, M.~A. (1989).
\newblock Supply function equilibria in oligopoly under uncertainty.
\newblock {\em Econometrica}, 57(6):1243--77.

\bibitem[\protect\astroncite{Luss}{1982}]{Luss82}
Luss, H. (1982).
\newblock Operations research and capacity expansion problems: A survey.
\newblock {\em Operations Research}, 30:907--947.

\bibitem[\protect\astroncite{Mart\'{\i}nez-de Alb\'{e}niz and
  Simchi-Levi}{2005}]{martinez2005}
Mart\'{\i}nez-de Alb\'{e}niz, V. and Simchi-Levi, D. (2005).
\newblock A portfolio approach to procurement contracts.
\newblock {\em Production and Operations Management}, 14(1):90--114.

\bibitem[\protect\astroncite{Mart\'{\i}nez-de Alb\'{e}niz and
  Simchi-Levi}{2009}]{MartinzeSimchi2009}
Mart\'{\i}nez-de Alb\'{e}niz, V. and Simchi-Levi, D. (2009).
\newblock Competition in the supply option market.
\newblock {\em Operations Research}, 57(5):1082--1097.

\bibitem[\protect\astroncite{Menezes and Monteiro}{1995}]{Menezes1995}
Menezes, F.~M. and Monteiro, P.~K. (1995).
\newblock Existence of equilibrium in a discriminatory price auction.
\newblock {\em Mathematical Social Sciences}, 30(3):285--292.

\bibitem[\protect\astroncite{Murota}{2003}]{Murota2003}
Murota, K. (2003).
\newblock {\em Discrete Convex Analysis}.
\newblock SIAM.

\bibitem[\protect\astroncite{Murota}{2009}]{Murota2009}
Murota, K. (2009).
\newblock Recent developments in discrete convex analysis.
\newblock In Cook, W., Lov\'{a}sz, L., and Vygen, J., editors, {\em Research
  Trends in Combinatorial Optimization}, chapter~11, pages 219--260. Springer
  Berlin Heidelberg.

\bibitem[\protect\astroncite{Murota and Shioura}{2004}]{Murota_Shioura2004}
Murota, K. and Shioura, A. (2004).
\newblock Conjugacy relationship between m-convex and l-convex functions in
  continuous variables.
\newblock {\em Mathematical Programming}, 101(3):415--433.

\bibitem[\protect\astroncite{Pei et~al.}{2011}]{Pei2011}
Pei, P. P.-E., Simchi-Levi, D., and Tunca, T.~I. (2011).
\newblock Sourcing flexibility, spot trading, and procurement contract
  structure.
\newblock {\em Operations Research}, 59(3):578--601.

\bibitem[\protect\astroncite{Rothkopf}{2007}]{Rothkopf07}
Rothkopf, M. (2007).
\newblock Thirteen reasons why the {Vickrey-Clarke-Groves} process is not
  practical.
\newblock {\em Operations Research}, 55(2):191--197.

\bibitem[\protect\astroncite{Rothkopf et~al.}{1990}]{Rothkopf1990}
Rothkopf, M., Teisberg, T., and Kahn, E. (1990).
\newblock Why are {V}ickrey auctions rare?
\newblock {\em Journal of Political Economy}, 98:94--109.

\bibitem[\protect\astroncite{Schummer and Vohra}{2003}]{Schummer2003}
Schummer, J. and Vohra, R. (2003).
\newblock Auctions for procuring options.
\newblock {\em Operations Research}, 51(1):41--51.

\bibitem[\protect\astroncite{Secomandi and Wang}{2012}]{Secomandi2012}
Secomandi, N. and Wang, M.~X. (2012).
\newblock A computational approach to the real option management of network
  contracts for natural gas pipeline transport capacity.
\newblock {\em Manufacturing \& Service Operations Management}, 14(3):441--454.

\bibitem[\protect\astroncite{Van~Mieghem}{2003}]{van-Mieghem03}
Van~Mieghem, J.~A. (2003).
\newblock Commissioned paper: Capacity management, investment, and hedging:
  Review and recent developments.
\newblock {\em Manufacturing and Service Operations Management}, 5(4):269--302.

\bibitem[\protect\astroncite{Vickrey}{1961}]{Vickrey1961}
Vickrey, W. (1961).
\newblock Counterspeculation, auctions, and competitive sealed tenders.
\newblock {\em Journal of Finance}, 16:8--37.

\bibitem[\protect\astroncite{Vives}{2011}]{vives2011}
Vives, X. (2011).
\newblock Strategic supply function competition with private information.
\newblock {\em Econometrica}, 79(6):1919--1966.

\bibitem[\protect\astroncite{Wilson}{1979}]{wilson1979}
Wilson, R. (1979).
\newblock Auctions of shares.
\newblock {\em The Quarterly Journal of Economics}, 93(4):675--689.

\bibitem[\protect\astroncite{Wu et~al.}{2005a}]{WuErkoc2005}
Wu, D., Erkoc, M., and Karabuk, S. (2005a).
\newblock Managing capacity in the high-tech industry: A review of literature.
\newblock {\em The Engineering Economist}, 50(2):125--158.

\bibitem[\protect\astroncite{Wu and Kleindorfer}{2005}]{Wu2005}
Wu, D.~J. and Kleindorfer, P.~R. (2005).
\newblock Competitive options, supply contracting, and electronic markets.
\newblock {\em Management Science}, 51(3):452--466.

\bibitem[\protect\astroncite{Wu et~al.}{2005b}]{Wudss2005}
Wu, D.~J., Kleindorfer, P.~R., and Sun, Y. (2005b).
\newblock {Optimal capacity expansion in the presence of capacity options}.
\newblock {\em Decision Support Systems}, 40(3-4):553--561.

\bibitem[\protect\astroncite{Wu et~al.}{2002}]{Wu2002}
Wu, D.~J., Kleindorfer, P.~R., and Zhang, J.~E. (2002).
\newblock Optimal bidding and contracting strategies for capital-intensive
  goods.
\newblock {\em European Journal of Operational Research}, 137(3):657--676.

\end{thebibliography}

\newpage

\renewcommand{\thesection}{A} \setcounter{equation}{0} \renewcommand{\theequation}{A-\arabic{equation}}

\section*{Appendix}

In this section, we prove the results derived in this paper and also give some basic definitions from \citet{Murota2003} for
Section~\ref{sec:submodularity}.

\subsection{Some definitions}
The set of real numbers is denoted by
$\mathbb{R}$, and $\bar{\mathbb{R}}=\mathbb{R}\cup\{+\infty\}$ and
$\underline{\mathbb{R}}= \mathbb{R}\cup\{-\infty\}$. The set of integers is
denoted by $\mathbb{Z}$, and $\bar{\mathbb{Z}}=\mathbb{Z}\cup\{+\infty\}$ and
$\underline{\mathbb{Z}}=\mathbb{Z}\cup\{-\infty\}$. We use $\mathbb{D}$ to
denote \emph{either} $\mathbb{Z}$ \emph{or} $\mathbb{R}$. Denote
$[n]=\{1,\ldots,n\}$ for any positive number $n$. The characteristic vector of
$S\subseteq[n]$ is denoted by $\chi_{S}\in\{0,1\}^{n}$. For $i\in[n]$, we
write $\chi_{i}$ for $\chi_{\{i\}}$, which is the $i$th unit vector, and
$\chi_{0}=\mathbf{0}$ (zero vector).

\subsubsection*{$M^{\natural}$-convexity}

For a function $f$: $\mathbb{D}^{n} \rightarrow\mathbb{R}\cup\{-\infty,
+\infty\}$, the set
\[
\text{dom}_{\mathbb{D}} f = \{x\in\mathbb{D}^{n}: f(x)\in\mathbb{R}\}
\]
is called the \emph{effective domain} of $f$. For a vector $z\in\mathbb{R}^{n}
$, define the \emph{positive} and \emph{negative supports} of $z$ as
\[
\text{supp}^{+}(z)=\{i\in[n]: z_{i} > 0\}, \quad\text{supp}^{-}(z)=\{i\in[n]:
z_{i} < 0\}.
\]
A function $f$: $\mathbb{Z}^{n}\rightarrow\bar{\mathbb{R}}$ is said
$M^{\natural}$-\emph{convex} if for any $x,y\in\text{dom}_{\mathbb{Z}}f$ and
any $i\in\text{supp}^{+}(x-y)$, there exists $j\in\text{supp}^{-}%
(x-y)\cup\{0\} $ such that the following \emph{exchange property} is
satisfied:
\[
f(x)+f(y) \ge f(x-\chi_{i} + \chi_{j}) + f(y+\chi_{i} -\chi_{j}).
\]
Similarly, a function $f$: $\mathbb{R}^{n}\rightarrow\bar{\mathbb{R}}$ is said
$M^{\natural}$-\emph{convex} if for any $x,y\in\text{dom}_{\mathbb{R}}f$ and
any $i\in\text{supp}^{+}(x-y)$, there exist $j\in\text{supp}^{-}%
(x-y)\cup\{0\}$ and $\lambda_{0} > 0$ such that
\[
f(x)+f(y) \ge f(x-\lambda(\chi_{i} - \chi_{j})) + f(y+\lambda(\chi_{i}
-\chi_{j}))
\]
for all $\lambda\in\mathbb{R}^{n}$ with $0\le\lambda\le\lambda_{0}$. A
function $f$: $\mathbb{D}^{n}\rightarrow\underline{\mathbb{R}}$ is said
$M^{\natural}$-\emph{concave} if $(-f)$ is $M^{\natural}$-convex.

\subsubsection*{Laminar convexity}

A non-empty set $\mathcal{L}\subseteq2^{[n]}$ is called a \emph{laminar
family} if for any $A,B\in\mathcal{L}$, we have $A\cap B = \emptyset$, or
$A\subseteq B$, or $B\subseteq A$. A function $f$: $\mathbb{D}^{n}
\rightarrow\bar{\mathbb{R}}$ is said \emph{laminar convex} if it can be
represented as
\[
f(x) = \sum_{S\in\mathcal{L}} f_{S}(x(S)),
\]
where $\{f_{S}\}$ are univariate convex functions, $\mathcal{L}$ is a laminar
family, and $x(S)=\sum_{i\in S} x_{i}$. A function $f$: $\mathbb{D}^{n}
\rightarrow\underline{\mathbb{R}}$ is said \emph{laminar concave} if $(-f)$ is
laminar convex.

\subsection{Proof of Theorem~\ref{thm:best-resp}}

(a) We consider any feasible offer $(\tilde{P}_{i}(\mathbf{s}),\tilde{R}_{i}(\mathbf{t}))$
from supplier $i$, giving a set of bids
\[
\tilde{\mathcal{B}}_{i}=\{(P_{j}(\mathbf{s}),R_{j}(\mathbf{t})): j\in N,
j\neq i\}\cup\{(\tilde{P}_{i}(\mathbf{s}),\tilde{R}_{i}(\mathbf{t}))\}.
\]
The buyer can obtain a profit of $\Pi_{\mathcal{B}_{i}}^{\ast}(N\backslash
\{i\})$ through restricting consideration to choices with $t_{i}=0$ (which
also implies $s_{i}=0$), and so we have a lower bound on buyer profit:
\begin{equation}\label{ineq:A1}
\Pi_{\tilde{\mathcal{B}}_{i}}^{\ast}(N)\geq\Pi_{\mathcal{B}_{i}}^{\ast}(N\backslash\{i\}).
\end{equation}
Taking $(\tilde{\mathbf{t}},\tilde{\mathbf{s}})$ as an optimal choice
by the buyer given bids $\tilde{\mathcal{B}}_{i}$, we have
\[
\Pi_{\tilde{\mathcal{B}}_{i}}^{\ast}(N)=\mathbb{E}\left[V(D,\tilde{\mathbf{s}}(D))
 -\sum_{j\neq i}(P_{j}(\tilde{\mathbf{s}}(D))+R_{j}(\tilde{\mathbf{t}}))
 -(\tilde{P}_{i}(\tilde{\mathbf{s}}(D))+\tilde{R}_{i}(\tilde{\mathbf{t}}))\right].
\]
Hence
\[
\mathbb{E}\left[\tilde{P}_{i}(\tilde{\mathbf{s}}(D))+\tilde{R}_{i}(\tilde{\mathbf{t}}))\right]
=\mathbb{E}\left[V(D,\tilde{\mathbf{s}}(D))-\sum_{j\neq i}(P_{j}(\tilde{\mathbf{s}}(D))
+R_{j}(\tilde{\mathbf{t}}))\right]-\Pi_{\tilde{\mathcal{B}}_{i}}^{\ast}(N).
\]
Using $\Pi_{\mathcal{B}_{i}}(\tilde{\mathbf{t}},\tilde{\mathbf{s}})\leq\Pi_{\mathcal{B}_{i}}^{\ast}(N)$
and \eqref{ineq:A1}, with a complete set of bids $\tilde{\mathcal{B}}_{i}$, we can calculate the profit for
supplier $i$ as follows:
\begin{align*}
\mathbb{E} & \left[\tilde{R}_{i}(\tilde{\mathbf{t}})  -E_{i}(\tilde{\mathbf{t}})+\tilde{P}_{i}(\tilde{\mathbf{s}}(D))
  -C_{i}(\tilde{\mathbf{s}}(D))\right] \\
       & =\mathbb{E}\left[V(D,\tilde{\mathbf{s}}(D))-\sum_{j\neq i}(P_{j}(\tilde{\mathbf{s}}(D))
  +R_{j}(\tilde{\mathbf{t}}))-E_{i}(\tilde{\mathbf{t}})-C_{i}(\tilde{\mathbf{s}}(D))\right]
  -\Pi_{\tilde{\mathcal{B}}_{i}}^{\ast}(N) \\
       & =\Pi_{\mathcal{B}_{i}}(\tilde{\mathbf{t}},\tilde{\mathbf{s}})-\Pi_{\tilde{\mathcal{B}}_{i}}^{\ast}(N)
  \leq\Pi_{\mathcal{B}_{i}}^{\ast}(N)-\Pi_{\mathcal{B}_{i}}^{\ast}(N\backslash\{i\}) = Z_{i},
\end{align*}
as required.

(b) Since $Z_{i}>0$, we know that $\Pi_{\mathcal{B}_{i}}^{\ast}(N)>
\Pi_{\mathcal{B}_{i}}^{\ast}(N\backslash\{i\})$, which implies that the optimal
choice $(\bar{\mathbf{t}},\bar{\mathbf{s}})$ for the buyer given
bids $\mathcal{B}_{i}$ has $\bar{t}_{i}>0$, otherwise these two would be
the same. We need to show that when the offer of $(\bar{P}_{i}(\cdot),\bar{R}_{i}(\cdot))$
is made the buyer will choose supplier $i$ (i.e., have $t_{i}>0$), which
will then automatically give a profit of $Z_{i}$ for supplier $i$ from \eqref{eq:supplier-profit}.
We will show that $(\bar{\mathbf{t}},\bar{\mathbf{s}})$ is optimal for the buyer given bids
\[
\bar{\mathcal{B}}_{i}=\{(P_{j}(\mathbf{s}),R_{j}(\mathbf{t})): j\in
N,j\neq i\}\cup\{(\bar{P}_{i}(\mathbf{s}),\bar{R}_{i}(\mathbf{t}))\},
\]
and this suffices since $i$ is preferred. Now by definition
\[
\Pi_{\bar{\mathcal{B}}_{i}}^{\ast}(N) =
 \max_{\substack{\mathbf{t}\in\mathbf{T} \\ \mathbf{s}(D)\in X(\mathbf{t})}}
 \mathbb{E}\left[  V(D,\mathbf{s}(D))-\sum_{j\neq i}(P_{j}(\mathbf{s}(D))+R_{j}(\mathbf{t}))-
 (\bar{P}_{i}(\mathbf{s}(D))+\bar{R}_{i}(\mathbf{t}))\right].
\]
Now
\[
\Pi_{\bar{\mathcal{B}}_{i}}(\bar{\mathbf{t}},\bar{\mathbf{s}})
=\Pi_{\mathcal{B}_{i}}(\bar{\mathbf{t}},\bar{\mathbf{s}})-Z_{i}
=\Pi_{\mathcal{B}_{i}}^{\ast}(N)-Z_{i},
\]
since $\bar{t}_{i}>0$. Now consider an arbitrary choice for the buyer
$(\mathbf{t},\mathbf{s})$ with $t_{i}>0$. This has
\[
\Pi_{\bar{\mathcal{B}}_{i}}(\mathbf{t},\mathbf{s})=
\Pi_{\mathcal{B}_{i}}(\mathbf{t},\mathbf{s})-Z_{i}\leq\Pi_{\mathcal{B}_{i}}^{\ast}(N)-Z_{i},
\]
and, moreover, any choice $(\mathbf{t},\mathbf{s})$ with $t_{i}=0$ has
\[
\Pi_{\bar{\mathcal{B}}_{i}}(\mathbf{t},\mathbf{s})=
\Pi_{\mathcal{B}_{i}}(\mathbf{t},\mathbf{s})\leq
\Pi_{\mathcal{B}_{i}}^{\ast}(N\backslash\{i\})=\Pi_{\mathcal{B}_{i}}^{\ast}(N)-Z_{i}.
\]
Thus we have shown the optimality we needed.

(c) In this case with the bid $(C_{i}(\mathbf{s}),\bar{R}_{i}(\mathbf{t})-\varepsilon)$ the profit for supplier $i$
is $Z_{i}-\varepsilon$ provided $t_{i}>0$. Given this offer, the buyer by choosing $(\bar{\mathbf{t}},
\bar{\mathbf{s}})$, defined in part (b), will achieve a profit of
$\Pi_{\mathcal{B}_{i}}^{\ast}(N)-Z_{i}+\varepsilon=\Pi_{\mathcal{B}_{i}}^{\ast}(N\backslash\{i\})+\varepsilon$,
which is therefore greater than any buyer profit available when $t_{i}=0$. Hence the buyer's
optimal choice must have $t_{i}>0$ and this completes the proof.

\subsection{Proof of Lemma~\ref{le-sub1}}

We prove the lemma by induction. It is trivial when $|N \setminus S| \le 1$, so we
begin with $|N \setminus S| = 2$, and we suppose that $N = S \cup \{j,k\}$.
From (\ref{eq-sub}), we obtain
\[
{\Pi} _{\mathcal{C}}^*(N) + {\Pi} _{\mathcal{C}}^*(N \setminus \{j,k\})
\le {\Pi} _{\mathcal{C}}^*(N \setminus \{j\})
+ {\Pi} _{\mathcal{C}}^*(N \setminus \{k\}),
\]
which can be rearranged to obtain the result required. Suppose (\ref{eq-sub-2}) holds for any $S$ with
$|N\setminus S|=k$. Then for $j\in S$, we have
\begin{align*}
\sum_{i\in N\setminus (S\cup \{j\})}
 & \left( {\Pi }_{\mathcal{C}}^{\ast }(N)-{\Pi }_{\mathcal{C}}^{\ast }(N\setminus \{i\})\right) \\
 & =\sum_{i\in N\setminus S}\left( {\Pi }_{\mathcal{C}}^{\ast }(N)-{\Pi }_{\mathcal{C}}^{\ast }
   (N\setminus \{i\})\right) +{\Pi }_{\mathcal{C}}^{\ast }(N)-{\Pi }_{\mathcal{C}}^{\ast }(N\setminus\{j\}) \\
 &\leq {\Pi }_{\mathcal{C}}^{\ast }(N)-{\Pi }_{\mathcal{C}}^{\ast }(S)+{\Pi }_{\mathcal{C}}^{\ast }(N)
  -{\Pi }_{\mathcal{C}}^{\ast }(N\setminus \{j\}) \\
 &\leq 2{\Pi }_{\mathcal{C}}^{\ast }(N)-({\Pi }_{\mathcal{C}}^{\ast }(N)
  +{\Pi }_{\mathcal{C}}^{\ast }(S\setminus \{j\}))
 = {\Pi }_{\mathcal{C}}^{\ast }(N)-{\Pi }_{\mathcal{C}}^{\ast }(S\setminus\{j\}),
\end{align*}
where the first inequality follows from the inductive hypothesis and the
second inequality follows from the submodularity property as assumed. Hence,
we establish that the result holds for $|N\setminus S|=k+1$, which completes
the proof.

\subsection{Proof of Lemma~\ref{le-sub2}}

We prove the lemma by contradiction. Suppose otherwise and there exists $j\in N^{\ast }(\mathcal{C})$ and $j\neq i$
such that $j\notin N_{-i}^{\ast }(\mathcal{C})$. By definition, $\Pi _{\mathcal{C}}^{\ast }(N\setminus\{i\})
=\Pi _{\mathcal{C}}^{\ast }(N\setminus \{i,j\})$. Moreover, inequality (\ref{eq-sub}) implies
$\Pi _{\mathcal{C}}^{\ast }(N\setminus \{i\})+\Pi _{\mathcal{C}}^{\ast }(N\setminus \{j\})\geq
\Pi _{\mathcal{C}}^{\ast}(N\setminus \{i,j\})+\Pi _{\mathcal{C}}^{\ast }(N)$. Thus, $\Pi _{\mathcal{C}}^{\ast }
(N\setminus \{j\})\geq \Pi _{\mathcal{C}}^{\ast }(N)$, which
contradicts the fact that $j\in N^{\ast }(\mathcal{C})$.

\subsection{Proof of Theorem~\ref{thm:general-equilib}}

We establish that with these bids the profit for supplier $i$ is
$\Pi_{\mathcal{C}}^{\ast}(N)-\Pi_{\mathcal{C}}^{\ast}(N\backslash\{i\})$, and
then that given the bids of the other players, no improvement on this is
possible for supplier $i$. We consider two cases. First suppose that
$\Pi_{\mathcal{C}}^{\ast}(N)-\Pi_{\mathcal{C}}^{\ast}(N\backslash\{i\})=0$
then $i\notin N^{\ast}(\mathcal{C})$ and with bids $\bar{\mathcal{B}}$
supplier $i$ makes zero profit. In the second case $i\in N^{\ast}(\mathcal{C})$ and
$\Pi_{\mathcal{C}}^{\ast}(N)>\Pi_{\mathcal{C}}^{\ast}(N\backslash\{i\})$. This implies that
any optimal solution for $\Pi_{\mathcal{C}}^{\ast}(N)$ must have $t_{i}>0$.

Next we show that the buyer has an optimal choice $(\mathbf{t}_{N}^{\ast},\mathbf{s}_{N}^{\ast})$
when facing bids $\bar{\mathcal{B}}$. Now for any feasible $(\mathbf{t},\mathbf{s})$,
\begin{align*}
\Pi_{\bar{\mathcal{B}}}(\mathbf{t},\mathbf{s})
 &  =\mathbb{E}[V(D,\mathbf{s}(D))-\sum_{i\in N^{\ast}(\mathcal{C})}(\bar{P}_{i}(\mathbf{s}(D))
    +\bar{R}_{i}(\mathbf{t})) \\
 & \quad -\sum_{i\notin N^{\ast}(\mathcal{C})}(C_{i}(\mathbf{s}(D))+E_{i}(\mathbf{t}))]\\
 &  =\mathbb{E}[V(D,\mathbf{s}(D))-\sum_{j\in N^{\ast}(\mathcal{C}),t_{j}>0}(\Pi_{\mathcal{C}}^{\ast}(N)
    -\Pi_{\mathcal{C}}^{\ast}(N\backslash\{j\})-\Delta_{j}(\mathbf{t})) \\
 &  \quad -\sum_{i\in N}(C_{i}(\mathbf{s}(D))+E_{i}(\mathbf{t}))]\\
 &  =\mathbb{E}[V(D,\mathbf{s}(D))+\sum_{j\in N^{\ast}(\mathcal{C}),t_{j}>0}\Delta_{j}(\mathbf{t})
    -\sum_{i\in N}(C_{i}(\mathbf{s}(D))+E_{i}(\mathbf{t}))] \\
 &  \quad -\sum_{j\in N}(\Pi_{\mathcal{C}}^{\ast}(N)-\Pi_{\mathcal{C}}^{\ast}(N\backslash\{j\}))
    +\sum_{j\in N\backslash I(\mathbf{t})}(\Pi_{\mathcal{C}}^{\ast}(N)-\Pi_{\mathcal{C}}^{\ast}(N\backslash\{j\}))\\
 &  \leq\mathbb{E}[V(D,\mathbf{s}(D))+\sum_{j\in N^{\ast}(\mathcal{C}),t_{j}>0}\Delta_{j}(\mathbf{t})
    -\sum_{i\in N}(C_{i}(\mathbf{s}(D))+E_{i}(\mathbf{t}))]\\
 &  \quad -\sum_{j\in N}(\Pi_{\mathcal{C}}^{\ast}(N)-\Pi_{\mathcal{C}}^{\ast}(N\backslash\{j\}))
    +\Pi_{\mathcal{C}}^{\ast}(N)-\Pi_{\mathcal{C}}^{\ast}(I(\mathbf{t})),
\end{align*}
where the inequality follows from Lemma~\ref{le-sub1} and we have used
$\Pi_{\mathcal{C}}^{\ast}(N)-\Pi_{\mathcal{C}}^{\ast}(N\backslash\{j\})=0$ for
$j\notin N^{\ast}(\mathcal{C})$. Since $\Delta_{i}(\mathbf{t})\geq0$, using
(\ref{ineq:consistent-variations}) we obtain
\begin{align*}
\sum_{j\in N^{\ast}(\mathcal{C}),\,t_{j}>0}\Delta_{j}(\mathbf{t})
 &  \leq \sum_{j\in I(\mathbf{t})}\Delta_{j}(\mathbf{t})
   \leq\Pi_{\mathcal{C}}^{\ast}(I(\mathbf{t}))-\Pi_{\mathcal{C}}(\mathbf{t},\mathbf{s})\\
 &  =\Pi_{\mathcal{C}}^{\ast}(I(\mathbf{t}))-\mathbb{E}[V(D,\mathbf{s}(D))
    -\sum_{i\in N}(C_{i}(\mathbf{s}(D))+E_{i}(\mathbf{t}))].
\end{align*}
Substitution into the inequality for $\Pi_{\bar{\mathcal{B}}}(\mathbf{t},\mathbf{s})$ leads to
\begin{equation}\label{ineq:B bar bound}
\Pi_{\bar{\mathcal{B}}}(\mathbf{t},\mathbf{s})\leq\Pi_{\mathcal{C}}^{\ast}(N)
 -\sum_{j\in N}(\Pi_{\mathcal{C}}^{\ast}(N)-\Pi_{\mathcal{C}}^{\ast}(N\backslash\{j\})).
\end{equation}

We will show that with the choice $(\mathbf{t}_{N}^{\ast},\mathbf{s}_{N}^{\ast})$ the buyer
can achieve this bound. Since $\sum_{j\in N}\Delta_{j}(\mathbf{t}_{N}^{\ast})=0$,
we have
\begin{align*}
\Pi_{\bar{\mathcal{B}}}(\mathbf{t}_{N}^{\ast},\mathbf{s}_{N}^{\ast})
&  =\mathbb{E}[V(D,\mathbf{s}_{N}^{\ast}(D))-\sum_{j\in N}(\bar{P}_{j}(\mathbf{s}_{N}^{\ast}(D))
   +\bar{R}_{j}(\mathbf{t}_{N}^{\ast}))]\\
&  =\mathbb{E}[V(D,\mathbf{s}_{N}^{\ast}(D))-\sum_{j\in N}(C_{j}(\mathbf{s}_{N}^{\ast}(D))+E_{j}(\mathbf{t}_{N}^{\ast})) \\
& \quad -\sum_{j\in N^{\ast}(\mathcal{C})}\left(  \Pi_{\mathcal{C}}^{\ast}(N)
  -\Pi_{\mathcal{C}}^{\ast}(N\backslash\{j\})\right)]\\
&  =\Pi_{\mathcal{C}}^{\ast}(N)-\sum_{j\in N}\left(\Pi_{\mathcal{C}}^{\ast}(N)
  -\Pi_{\mathcal{C}}^{\ast}(N\backslash\{j\})\right).
\end{align*}
Here we have used the fact that $N^{\ast}(\mathcal{C})= I(\mathbf{t}%
_{N}^{\ast})$ and also that $\Pi_{\mathcal{C}}^{\ast}(N)-\Pi_{\mathcal{C}%
}^{\ast}(N\backslash\{j\})=0$ when $j\notin N^{\ast}(\mathcal{C})$. Hence we
have established that $(\mathbf{t}_{N}^{\ast},\mathbf{s}_{N}^{\ast})$ is an
optimal choice for the buyer. By Assumption~\ref{assum-SCopt} the buyer choice
$(\mathbf{t}_{N}^{\ast},\mathbf{s}_{N}^{\ast})$ that achieves supply chain
optimality is preferred and hence we have $t_{i}>0$ , from which it follows
that the profit for supplier $i$ when bids are $\bar{\mathcal{B}}$ is
$\Pi_{C}^{\ast}(N)-\Pi_{\mathcal{C}}^{\ast}(N\backslash\{i\})$.
Now we evaluate $\Pi_{\bar{\mathcal{B}}}^{\ast}(N\backslash\{i\})$. For
any feasible $(\mathbf{t},\mathbf{s})$, the bound (\ref{ineq:B bar bound})
still applies and hence
\[
\Pi_{\bar{\mathcal{B}}}(\mathbf{t},\mathbf{s})\leq\sum_{j\in N}\Pi
_{C}^{\ast}(N\backslash\{j\})-(\left\vert N\right\vert -1)\Pi_{\mathcal{C}%
}^{\ast}(N).
\]
Since $\sum_{j\in N\backslash\{i\}}\Delta_{j}(\mathbf{t}_{N\backslash
\{i\}}^{\ast})=0$ and each $\Delta_{j}$ is non-negative we have $\Delta
_{j}(\mathbf{t}_{N\backslash\{i\}}^{\ast})=0$ for each $j\in N\backslash
\{i\}$. Thus
\begin{align*}
\Pi_{\bar{\mathcal{B}}}(\mathbf{t}_{N\backslash\{i\}}^{\ast},\mathbf{s}_{N\backslash\{i\}}^{\ast})
 &  =\mathbb{E}[V(D,\mathbf{s}_{N\backslash\{i\}}^{\ast}(D))
      -\sum_{j\in N^{\ast}(\mathcal{C})}(\bar{P}_{j}(\mathbf{s}_{N\backslash\{i\}}^{\ast}(D))
     +\bar{R}_{j}(\mathbf{t}_{N\backslash\{i\}}^{\ast})) \\
 &  \quad -\sum_{j\notin N^{\ast}(\mathcal{C})}(C_{j}(\mathbf{s}_{N\backslash\{i\}}^{\ast}(D))
    +E_{j}(\mathbf{t}_{N\backslash\{i\}}^{\ast}))] \\
 &  =\mathbb{E}[V(D,\mathbf{s}_{N\backslash\{i\}}^{\ast}(D))-\sum_{j\in N}(C_{j}(\mathbf{s}_{N\backslash\{i\}}^{\ast}(D))
    +E_{j}(\mathbf{t}_{N\backslash\{i\}}^{\ast}))] \\
 & \quad -\sum_{j\in I(\mathbf{t}_{N\backslash\{i\}}^{\ast})\cap N^{\ast}(\mathcal{C})}\left(\Pi_{\mathcal{C}}^{\ast}(N)
   -\Pi_{\mathcal{C}}^{\ast}(N\backslash\{j\})\right) \\
 &  =\Pi_{\mathcal{C}}^{\ast}(N\backslash\{i\})-\sum_{j\in N^{\ast}(\mathcal{C})\backslash\{i\}}
   \left(  \Pi_{\mathcal{C}}^{\ast}(N)-\Pi_{\mathcal{C}}^{\ast}(N\backslash\{j\})\right),
\end{align*}
where the last equality is based on Lemma~\ref{le-sub2}. Hence, as $\Pi_{\mathcal{C}}^{\ast}(N)=
\Pi_{\mathcal{C}}^{\ast}(N\backslash\{j\}$ for $j\notin N^{\ast}(\mathcal{C})$, we have
\begin{align*}
\Pi_{\bar{\mathcal{B}}}(\mathbf{t}_{N\backslash\{i\}}^{\ast}%
,\mathbf{s}_{N\backslash\{i\}}^{\ast})  &  =\Pi_{\mathcal{C}}^{\ast
}(N\backslash\{i\})-\sum_{j\in N,j\neq i}\left(  \Pi_{\mathcal{C}}^{\ast
}(N)-\Pi_{\mathcal{C}}^{\ast}(N\backslash\{j\})\right) \\
&  =\sum_{j\in N}\Pi_{\mathcal{C}}^{\ast}(N\backslash\{j\})-(\left\vert
N\right\vert -1)\Pi_{\mathcal{C}}^{\ast}(N) ,
\end{align*}
and so $\left(\mathbf{t}_{N\backslash\{i\}}^{\ast},\mathbf{s}_{N\backslash\{i\}}^{\ast}\right)$ is also optimal
for the buyer faced with bids $\bar{\mathcal{B}}$. Thus
\begin{equation}\label{eq:Bbar-values}
\Pi_{\bar{\mathcal{B}}}^{\ast}(N\backslash\{i\})=\Pi_{\bar
{\mathcal{B}}}^{\ast}(N)=\sum_{j\in N}\Pi_{\mathcal{C}}^{\ast}(N\backslash
\{j\})-(\left\vert N\right\vert -1)\Pi_{\mathcal{C}}^{\ast}(N),
\end{equation}
and there is no loss to the buyer from a restriction that one of the $t_{i}$
values is zero.

Now suppose that there is a different offer $(\tilde{P}_{i}(\mathbf{s}),\tilde{R}_{i}(\mathbf{t}))$
by one of the suppliers $i\in N^{\ast}(\mathcal{C})$, giving a set of bids
\[
\tilde{\mathcal{B}}_{i}=\{(\bar{P}_{j}(\mathbf{s}),\bar{R}_{j}(\mathbf{t})):
 i\neq j \in N^{\ast}(\mathcal{C})\}\cup\{(\tilde{P}_{i}(\mathbf{s}),\tilde{R}_{i}(\mathbf{t}))\}
 \cup\{(C_{j}(\mathbf{s}),E_{j}(\mathbf{t})): j \notin N^{\ast}(\mathcal{C})\}.
\]
The buyer can obtain a profit of $\Pi_{\bar{\mathcal{B}}}^{\ast
}(N\backslash\{i\})$ through restricting consideration to choices with
$t_{i}=0$. This follows because $\tilde{B}_{i}(\mathbf{t})=0$ when
$t_{i}=0$. We can write this lower bound on buyer profit as%
\[
\Pi_{\tilde{\mathcal{B}}_{i}}^{\ast}(N)\geq\Pi_{\bar{\mathcal{B}}%
}^{\ast}(N\backslash\{i\}).
\]
Suppose that $(\tilde{\mathbf{t}},\tilde{\mathbf{s}})$ is an
optimal choice by the buyer given bids $\tilde{\mathcal{B}}_{i}$, so
\begin{align*}
\Pi_{\tilde{\mathcal{B}}_{i}}^{\ast}(N)
 & =\mathbb{E}[V(D,\tilde{\mathbf{s}}(D))-\sum_{j\in N^{\ast}(\mathcal{C})\backslash\{i\}}
   (\bar{P}_{j}(\tilde{\mathbf{s}}(D))-\bar{R}_{j}(\tilde{\mathbf{t}})) \\
 & \quad -(\tilde{P}_{i}(\tilde{\mathbf{s}}(D))-\tilde{R}_{i}(\tilde{\mathbf{t}}))
   -\sum_{j\notin N^{\ast}(\mathcal{C})}(C_{j}(\tilde{\mathbf{s}}(D))-E_{j}(\tilde{\mathbf{t}}))].
\end{align*}
Given bids $\tilde{\mathcal{B}}_{i}$, the profit for supplier $i$ is
\begin{align*}
\mathbb{E}[\tilde{P}_{i}(\tilde{\mathbf{s}}(D)) &+\tilde{R}_{i}(\tilde{\mathbf{t}})-(C_{i}(\tilde{\mathbf{s}}(D))
  +E_{i}(\tilde{\mathbf{t}}))] \\
&  =\mathbb{E}[V(D,\tilde{\mathbf{s}}(D))-\sum_{j\in N^{\ast}(\mathcal{C})\backslash\{i\}}
  (\bar{P}_{j}(\tilde{\mathbf{s}}(D))+\bar{R}_{j}(\tilde{\mathbf{t}}))\\
& \quad  -\sum_{j\notin N^{\ast}(\mathcal{C})}(C_{j}(\tilde{\mathbf{s}}(D))+E_{j}(\tilde{\mathbf{t}}))
  -\Pi_{\tilde{\mathcal{B}}_{i}}^{\ast}(N)-(C_{i}(\tilde{\mathbf{s}}(D))+E_{i}(\tilde{\mathbf{t}}))] \\
&  =\Pi_{\mathcal{C}}(\tilde{\mathbf{t}},\tilde{\mathbf{s}})
   -\sum_{\substack{j\in N^{\ast}(\mathcal{C})\backslash\{i\}\\ \tilde{\mathbf{t}}_{j}>0}}
   \left(  \Pi_{\mathcal{C}}^{\ast}(N)-\Pi_{\mathcal{C}}^{\ast}(N\backslash\{j\})-\Delta_{j}(\tilde{\mathbf{t}})\right)
   -\Pi_{\tilde{\mathcal{B}}_{i}}^{\ast}(N).
\end{align*}
Since $\Delta_{j}(\tilde{\mathbf{t}})\geq0$ and
$\Pi_{\tilde{\mathcal{B}}_{i}}^{\ast}(N)\geq\Pi_{\bar{\mathcal{B}}}^{\ast}(N\backslash\{i\})$, the above quantity
is at most
\begin{align*}
\Pi_{\mathcal{C}}(\tilde{\mathbf{t}},\tilde{\mathbf{s}})
   & +\sum_{j\in I(\tilde{\mathbf{t}})}\Delta_{j}(\tilde{\mathbf{t}}) 
   -\sum_{\substack{j\in N^{\ast}(\mathcal{C})\backslash\{i\}\\\tilde
{\mathbf{t}}_{j}>0}}\left(  \Pi_{\mathcal{C}}^{\ast}(N)-\Pi_{\mathcal{C}}^{\ast}(N\backslash\{j\})\right)
-\Pi_{\bar{\mathcal{B}}}^{\ast}(N\backslash\{i\})\\
&  \leq\Pi_{\mathcal{C}}^{\ast}(I(\tilde{\mathbf{t}}))-\sum_{j\in N,j\neq
i}\left(  \Pi_{\mathcal{C}}^{\ast}(N)-\Pi_{\mathcal{C}}^{\ast}(N\backslash
\{j\})\right)  \\
& \quad +\sum_{j\notin I(\tilde{\mathbf{t}}),j\neq i}\left(
\Pi_{\mathcal{C}}^{\ast}(N)-\Pi_{\mathcal{C}}^{\ast}(N\backslash\{j\})\right)
-\Pi_{\bar{\mathcal{B}}}^{\ast}(N\backslash\{i\})\\
&  \leq\Pi_{\mathcal{C}}^{\ast}(I(\tilde{\mathbf{t}}))
   -\sum_{\substack{j\in N \\ j\neq i}}\left(\Pi_{\mathcal{C}}^{\ast}(N)-\Pi_{\mathcal{C}}^{\ast}(N\backslash\{j\})\right)  
   +\Pi_{\mathcal{C}}^{\ast}(N)-\Pi_{\mathcal{C}}^{\ast}(I(\tilde{\mathbf{t}}))
   -\Pi_{\bar{\mathcal{B}}}^{\ast}(N\backslash\{i\})\\
&  =\Pi_{\mathcal{C}}^{\ast}(N)-(\left\vert N\right\vert -1)\Pi_{\mathcal{C}}^{\ast}(N)
   +\sum_{j\neq i}\Pi_{\mathcal{C}}^{\ast}(N\backslash\{i\})-\Pi_{\bar{\mathcal{B}}}^{\ast}(N\backslash\{i\}),
\end{align*}
where the first inequality is from the fact that functions $\{\Delta_{i}(\mathbf{t}): i\in N\}$ are
consistent with the costs, while the second inequality is from Lemma~\ref{le-sub1}.
We can use (\ref{eq:Bbar-values}) and cancel terms to obtain
\[
\mathbb{E}[(\tilde{P}_{i}(\tilde{\mathbf{s}}(D))+\tilde{R}_{i}(\tilde{\mathbf{t}})
-(C_{i}(\tilde{\mathbf{s}}(D))+E_{i}(\tilde{\mathbf{t}}))]\leq\Pi_{\mathcal{C}}^{\ast}(N)
-\Pi_{\mathcal{C}}^{\ast}(N\backslash\{i\}).
\]

The final step is to consider an offer $(\tilde{P}_{i}(\mathbf{s}),\tilde{R}_{i}(\mathbf{t}))$
by one of the suppliers $i\notin N^{\ast}(\mathcal{C})$. For these suppliers there is no profit
under the set of bids $\bar{\mathcal{B}}$. We will show that there is no chance to make a
profit with a different bid. \ It is clear that any choice that has $t_{i}>0$
will give the buyer less than if the same choice was used with the bids
$\bar{\mathcal{B}}$, and we have seen that for $\bar{\mathcal{B}}$
an optimal choice is given by $(\mathbf{t}_{N}^{\ast},\mathbf{s}_{N}^{\ast})$. Thus this choice that
has $t_{i}=0$ (and gives preference to those in
the supply chain optimal choice) gives the buyer at least the same profit as
any other and is therefore chosen by the buyer. Hence the profit made by
supplier $i$ with the new offer can never be greater than zero.
Thus we have established that in both cases no other bid can achieve a higher
profit for supplier $i$, establishing that the bids given in
the theorem are indeed a Nash equilibrium.

\subsection{Proof of Corollary~\ref{Corr-br-n}}

This is immediate, since Theorem~\ref{thm:best-resp} part (a) establishes that supplier $i$
can make no more than $\Pi_{\mathcal{B}_{i}}^{\ast}(N)-\Pi_{\mathcal{B}_{i}}^{\ast}(N\backslash\{i\})$,
which by construction is the supplier $i$ profit
under the conditions of the Corollary. Moreover the example of an optimal
choice for the supplier corresponds to part (b) of Theorem~\ref{thm:best-resp}.
Finally, the statement on $\epsilon$-optimality follows from part (c) of Theorem~\ref{thm:best-resp}.

\subsection{Proof of Proposition~\ref{prop:equilibrium-power}}

First we want to show that $\theta_{i}>0$, for $i\in N^{\ast}(\mathcal{C})$.
Assumption~\ref{assum-SCopt} is enough (from uniqueness of the support of
supply chain capacity choices) to show $(\mathbf{t}_{N}^{\ast})_{i}>0$, as
we observed in (\ref{eq:support_tNstar}). Then use of submodularity and
repeated application of Lemma~\ref{le-sub2} with $N$ decreasing in size shows
that $(\mathbf{t}_{S}^{\ast})_{i}>0$ for $S\subseteq N$.

It suffices to show that the execution and reservation payments induced by
$p_{i}^{\ast}(s)$ and $r_{i}^{\ast}(t)$ satisfy the conditions given in
Theorem~\ref{thm:general-equilib}. It is straightforward that the execution
payment is $\int_{0}^{s_{i}}p_{i}^{\ast}(s)ds=\int_{0}^{s_{i}}c_{i}%
(s)ds=C_{i}(s_{i})$ for $i\in N$. For $i\notin N^{\ast}(\mathcal{C})$,
$\theta_{i}=0$ and thus $\int_{0}^{t_{i}}r_{i}^{\ast}(t)dt=E_{i}(t_{i})$. For
$i\in N^{\ast}(\mathcal{C})$, we obtain the reservation payment as follows:
\begin{align*}
\int_{0}^{t_{i}}r_{i}^{\ast}(t)dt &  =\int_{0}^{t_{i}}e_{i}(t)+\int_{0}%
^{\min(t_{i},\theta_{i})}\delta_{i}(t;\beta_{i})dt\\
&  =E_{i}(t_{i})+(\Pi_{\mathcal{C}}^{\ast}(N)-\Pi_{\mathcal{C}}^{\ast
}(N\setminus\{i\}))\int_{0}^{\min(t_{i},\theta_{i})}\frac{(\beta_{i}%
+1)}{\theta_{i}}\left(  \frac{\theta_{i}-t}{\theta_{i}}\right)  ^{\beta_{i}%
}dt\\
&  =E_{i}(t_{i})+(\Pi_{\mathcal{C}}^{\ast}(N)-\Pi_{\mathcal{C}}^{\ast
}(N\setminus\{i\}))\left(  1-\left(  \frac{\theta_{i}-\min(t_{i},\theta_{i}%
)}{\theta_{i}}\right)  ^{\beta_{i}+1}\right)  .
\end{align*}
Thus,
\[\Delta_{i}(t_{i};\beta_{i})=(\Pi_{\mathcal{C}}^{\ast}(N)-\Pi
_{\mathcal{C}}^{\ast}(N\setminus\{i\}))\left(  \frac{\theta_{i}-\min
(t_{i},\theta_{i})}{\theta_{i}}\right)  ^{\beta_{i}+1}\geq0.
\]

Next we show that when all $\beta_{i}$s are large enough, the set of
$\Delta_{i}(t_{i};\beta_{i})$ derived from the power function bids is
consistent with the costs. There are two requirements for consistency, first
we need $\sum_{j\in S}\Delta_{j}((\mathbf{t}_{S}^{\ast})_{j};\beta_{j})=0$.
\ This follows because $(\mathbf{t}_{S}^{\ast})_{j}\geq\theta_{j}$ for $j\in
S$, and hence $\Delta_{j}((\mathbf{t}_{S}^{\ast})_{j};\beta_{j})=0$
independent of the choice of $\beta_{j}$. The second requirement is that for
any $\mathbf{t}$ with $I(t)\subseteq S$,%
\[
\sum_{j\in S}\Delta_{j}(t_{j})\leq\Pi_{\mathcal{C}}^{\ast}(S)
-\Pi_{\mathcal{C}}(\mathbf{t}).
\]

The right hand side of this expression is also a sum over the non zero elements
of $\mathbf{t}$ from (\ref{eqn-supply-chain-profit-n}). Thus we can look at
this inequality on a component by component basis, and we want to find values
of $\beta$ large enough that, for every $\mathbf{t}$%
\[
(\Pi_{\mathcal{C}}^{\ast}(N)-\Pi_{\mathcal{C}}^{\ast}(N\setminus\{i\}))\left(
\frac{\theta_{i}-\min(t_{i},\theta_{i})}{\theta_{i}}\right)  ^{\beta_{i}%
+1}\leq W(\mathbf{t),}%
\]
where
\begin{align}\label{ineq:power-function-proof1}
W(\mathbf{t}) & =\int_{0}^{\left(\mathbf{t}_{S}^{\ast}\right)_{i}}\left\{[\rho-c_{i}(x)]\bar{F}(\bar{h}_{i}
 (x,\left(  \mathbf{t}_{S}^{\ast}\right)_{-i}))-e_{i}(x)\right\}dx \nonumber \\
& \quad -\int_{0}^{t_{i}}\left\{[\rho -c_{i}(x)]\bar{F}(\bar{h}_{i}(x,\left(\mathbf{t}\right)_{-i}))-e_{i}(x)\right\}dx,
\end{align}
and $\mathbf{t}_{S}^{\ast}$ is optimal for $\Pi_{\mathcal{C}}^{\ast}(S)$.
We note that
\[
\Pi_{\mathcal{C}}^{\ast}(N)-\Pi_{\mathcal{C}}^{\ast}(N\setminus\{i\})
\leq\Pi_{\mathcal{C}}^{\ast}(S)-\Pi_{\mathcal{C}}^{\ast}(S\setminus\{i\})
\leq\Pi_{\mathcal{C}}^{\ast}(S)-\Pi_{\mathcal{C}}(\bar{\mathbf{t}}_{S,i}^{\ast}),
\]
where $\bar{\mathbf{t}}_{S,i}^{\ast}=((\mathbf{t}_{S}^{\ast})_{1},(\mathbf{t}_{S}^{\ast})_{2},
\ldots, (\mathbf{t}_{S}^{\ast})_{i-1},0,(\mathbf{t}_{S}^{\ast})_{i+1},\ldots, (\mathbf{t}_{S}^{\ast})_{n})$, i.e., it
consists of $\mathbf{t}_{S}^{\ast}$ with the $i$'th component set to zero. The
first inequality here comes from submodularity and the second from the
optimality of $\Pi_{\mathcal{C}}^{\ast}(S\setminus\{i\})$. Now we observe that
for $j\neq i$ the buyer's profit from supplier $j$, which is
\[
\int_{0}^{\left(  \mathbf{t}_{S}^{\ast}\right)  _{j}}\left\{  [\rho
-c_{j}(x)]\bar{F}(\bar{h}_{j}(x,\left(  \mathbf{t}_{S}^{\ast}\right)
_{-j}))-e_{j}(x)\right\}  dx,
\]
is increased by taking out the capacity from supplier $i$ (i.e., moving from
$\mathbf{t}_{S}^{\ast}$ to $\bar{\mathbf{t}}_{S,i}^{\ast}$) since
$\bar{h}_{j}(x,\left(  \mathbf{t}_{S}^{\ast}\right)_{-j}$ is reduced. Hence
\begin{align*}
\Pi_{\mathcal{C}}^{\ast}(S) &  \leq\sum_{j\in S,j\neq i}\int_{0}%
^{(\bar{\mathbf{t}}_{S,i}^{\ast})_{j}}\left\{  [\rho-c_{j}(x)]\bar
{F}(\bar{h}_{j}(x,(\bar{\mathbf{t}}_{S,i}^{\ast})_{-j}))-e_{j}%
(x)\right\}  dx \\
& \quad +\int_{0}^{\left(  \mathbf{t}_{S}^{\ast}\right)  _{i}}\left\{
[\rho-c_{i}(x)]\bar{F}(\bar{h}_{i}(x,\left(  \mathbf{t}_{S}^{\ast}\right)
_{-i}))-e_{i}(x)\right\}  dx\\
&  =\Pi_{\mathcal{C}}(\bar{\mathbf{t}}_{S,i}^{\ast})+\int_{0}^{\left(
\mathbf{t}_{S}^{\ast}\right)  _{i}}\left\{  [\rho-c_{i}(x)]\bar{F}(\bar{h}%
_{i}(x,\left(  \mathbf{t}_{S}^{\ast}\right)  _{-i}))-e_{i}(x)\right\}  dx.
\end{align*}
Thus,
\[
(\Pi_{\mathcal{C}}^{\ast}(N)-\Pi_{\mathcal{C}}^{\ast}(N\setminus\{i\}))\left(
\frac{\theta_{i}-\min(t_{i},\theta_{i})}{\theta_{i}}\right)  ^{\beta_{i}%
+1}\leq W_{\beta_{i}}(\mathbf{t)},
\]
where
\[
W_{\beta_{i}}(\mathbf{t})=
\left(  \frac{\theta_{i}-\min(t_{i},\theta_{i})}{\theta_{i}}\right)^{\beta_{i}+1}
\int_{0}^{\left(  \mathbf{t}_{S}^{\ast}\right)_{i}}
\left\{[\rho-c_{i}(x)]\bar{F}(\bar{h}_{i}(x,\left(\mathbf{t}_{S}^{\ast}\right)_{-i}))-e_{i}(x)\right\}dx.
\]
We need to show that $W_{\beta_{i}}(\mathbf{t)\leq}W(\mathbf{t)}$.  Observe
from (\ref{ineq:power-function-proof1})  that $W_{\beta_{i}}(\mathbf{t)=}%
W(\mathbf{t)}$ if $t_{i}=0$. Moreover $W_{\beta_{i}}(\mathbf{t)}$ is zero when
$t_{i}=\theta_{i}$, and hence $W_{\beta_{i}}(\mathbf{t)\leq}W(\mathbf{t)}$ at
$t_{i}=\theta_{i}$, with equality in the case that $\theta_{i}=$ $\left(
\mathbf{t}_{S}^{\ast}\right)_{i}$.

Since $\left(  \mathbf{t}_{S}^{\ast}\right)  _{i}$ gives the optimal solution
for $\Pi_{\mathcal{C}}^{\ast}(S)$ we know that
\[
\frac{\partial}{\partial t_{i}}W(\mathbf{t}_{S}^{\ast})=\left(
\left(  \rho-c_{i}(\left(  \mathbf{t}_{S}^{\ast}\right)_{i})\right)
\bar{F}(\bar{h}_{i}(t_{i},\left(\mathbf{t}_{S}^{\ast}\right)_{-i}))
-e_{i}(\left(\mathbf{t}_{S}^{\ast}\right)_{i})\right)  =0.
\]
Looking at the second derivative:
\[
\frac{\partial^{2}}{\partial t_{i}^{2}}W(\mathbf{t)=}\frac{\partial}{\partial
t_{i}}\left(  \left(  \rho-c_{i}(t_{i})\right)  \bar{F}(\bar{h}_{i}%
(t_{i},\left(  \mathbf{t}\right)  _{-i}))-e_{i}(t_{i})\right)  <0
\]
from our assumptions on the functions $c_{i}$ and $e_{i}$. The strict
inequality here can be established from observing that $\bar{F}$ is decreasing
and $\bar{h}_{i}(t_{i},\left(  \mathbf{t}\right)  _{-i})$ is increasing in
$t_{i}$. Now for $t_{i}\leq\theta_{i}$
\begin{align*}
\frac{\partial}{\partial t_{i}}W_{\beta_{i}}(\mathbf{t)}  & \mathbf{=-}%
\frac{\mathbf{(}\beta_{i}+1)}{\theta_{i}}\left(  \frac{\theta_{i}-t_{i}%
}{\theta_{i}}\right)  ^{\beta_{i}}\int_{0}^{\left(  \mathbf{t}_{S}^{\ast
}\right)  _{i}}\left\{  [\rho-c_{i}(x)]\bar{F}(\bar{h}_{i}(x,\left(
\mathbf{t}_{S}^{\ast}\right)  _{-i}))-e_{i}(x)\right\}  dx,\\
\frac{\partial^{2}}{\partial t_{i}^{2}}W_{\beta_{i}}(\mathbf{t)}  &
\mathbf{=}\frac{\mathbf{(}\beta_{i}+1)\beta_{i}}{\theta_{i}^{2}}\left(
\frac{\theta_{i}-t_{i}}{\theta_{i}}\right)  ^{\beta_{i}-1}\int_{0}^{\left(
\mathbf{t}_{S}^{\ast}\right)  _{i}}\left\{  [\rho-c_{i}(x)]\bar{F}(\bar{h}%
_{i}(x,\left(  \mathbf{t}_{S}^{\ast}\right)  _{-i}))-e_{i}(x)\right\}  dx,
\end{align*}
which are both zero at $t_{i}=\theta_{i}$. Hence a\ Taylor series expansion
shows that for $t_{i}$ close enough to $\theta_{i}$ but below it, we will have
$W_{\beta_{i}}(\mathbf{t)<}$ $W(\mathbf{t)}$. The multiplier $\left(
\frac{\theta_{i}-\min(t_{i},\theta_{i})}{\theta_{i}}\right)  ^{\beta_{i}+1}$
decreases as $\beta_{i}$ increases and has limit zero for any $t_{i}>0$. Note
that $\frac{\partial}{\partial t_{i}}W_{\beta_{i}}(\mathbf{t})$ decreases
towards\textbf{ }$\mathbf{-\infty}$  at $t_{i}=0$ as $\beta_{i}$ increases,
and so is less than $\frac{\partial}{\partial t_{i}}W(\mathbf{t}_{S}^{\ast
}\mathbf{)}$ for $\beta_{i}$ large enough. Hence we can set $\beta_{i}$ large
enough that $W_{\beta_{i}}(\mathbf{t)<}$ $W(\mathbf{t)}$ for all $t_{i}%
\in(0,\theta_{i})$. Therefore, we have established that the set of $\Delta
_{i}(t_{i};\beta_{i})$ is consistent with the costs.

\subsection{Proof of Proposition~\ref{prop-subadditive}}

For submodularity we simply need to show that ${\Pi}_{\mathcal{C}}^{\ast}(\{1,2\})
\leq{\Pi}_{\mathcal{C}}^{\ast}(\{1\})+{\Pi}_{\mathcal{C}}^{\ast}(\{2\})$. Let
$(\mathbf{t}^{\ast},\mathbf{s}^{\ast}(D))$ be the optimal supply chain choice when both suppliers are available.
Then
\begin{align*}
{\Pi}_{\mathcal{C}}^{\ast}(\{1,2\})  &  =\mathbb{E}[V(D,\mathbf{s}^{\ast
}(D))-\sum_{i\in\{1,2\}}(C_{i}(\mathbf{s}^{\ast}(D))+E_{i}(\mathbf{t}^{\ast
}))]\\
&  =\mathbb{E}[V(D,[\mathbf{s}^{\ast}(D)]_{1}+[\mathbf{s}^{\ast}(D)]_{2}%
) \\
& \quad -\sum_{i\in\{1,2\}}(C_{i}([\mathbf{s}^{\ast}(D)]_{1}+[\mathbf{s}^{\ast
}(D)]_{2})+E_{i}([\mathbf{t}^{\ast}]_{1}+[\mathbf{t}^{\ast}]_{2}))],
\end{align*}
where for any vector $\mathbf{x}$ we define $[\mathbf{x}]_{1}:=(x_{1},0)$ and
$[\mathbf{x}]_{2}:=(0,x_{2})$. Using subadditivity of $V$ and the restriction
on $C$ and $E$ functions we have
\begin{align*}
{\Pi}_{\mathcal{C}}^{\ast}(\{1,2\})  &  \leq\mathbb{E}[V(D,[\mathbf{s}^{\ast
}(D)]_{1})+V(D,[\mathbf{s}^{\ast}(D)]_{2})-\sum_{i\in\{1,2\}}(C_{i}%
([\mathbf{s}^{\ast}(D)]_{i})+E_{i}([\mathbf{t}^{\ast}]_{i}))]\\
&  ={\Pi}_{\mathcal{C}}([\mathbf{t}^{\ast}]_{1},[\mathbf{s}^{\ast}%
(D)]_{1})+{\Pi}_{\mathcal{C}}([\mathbf{t}^{\ast}]_{2},[\mathbf{s}^{\ast}(D)]_{2})
 \leq{\Pi}_{\mathcal{C}}^{\ast}(\{1\})+{\Pi}_{\mathcal{C}}^{\ast}(\{2\}).
\end{align*}
The final stage is simply to observe that for the capacity game with function
bids, we have $V(D,\mathbf{s})=\rho\min(D,s_{1}+s_{2})$ and this is a
subadditive function, since
\begin{align*}
V(D,\mathbf{s}_{A}+\mathbf{s}_{B})&=\min(D,s_{A1}+s_{A2}+s_{B1}+s_{B2}) \\
                                  &\leq\min(D,s_{A1}+s_{A2})+\min(D,s_{B1}+s_{B2}).
\end{align*}

\subsection{Proof of Theorem~\ref{thm:submodular}}

Note that the supply chain optimal profit as a set function can be expressed as follows:
\[
\Pi^*(\sigma)=\max\left\{\Pi_{\mathcal{C}}(\mathbf{t}):
\mathbf{t}\le \sigma\bar{d},\right\}, \ \sigma\in\{0,1\}^n.
\]
According to \eqref{eqn:channel_profit}, with the conditions of the theorem, $\Pi_{\mathcal{C}}(\mathbf{t})$ is
laminar concave and hence $M^\natural$-concave in $\mathbf{t}\in \mathbf{T}$ (see the
Appendix for definition) according to \citet{Murota2003}, which implies that
$\tilde{\Pi}_{\mathcal{C}}(\mathbf{t}) = \Pi_{\mathcal{C}}(\mathbf{t}\bar{d})$ is $M^\natural$-concave in
$\mathbf{t}\in [0,1]^n$. For any $\bar{\sigma}\in [0,1]^n$, let
\[
\tilde{\Pi}(\bar{\sigma})=\max\left\{\tilde{\Pi}_{\mathcal{C}}(\mathbf{t}):
\mathbf{0}\le \mathbf{t}\le \bar{\sigma}\right\}.
\]
We show that $\tilde{\Pi}(\bar{\sigma})$ is $M^\natural$-concave over $[0,1]^n$. To this end,
consider the informal convolution function
\[
\bar{\Pi}(\bar{\sigma}) = \sup\left\{\bar{\Pi}_{\mathcal{C}}(\mathbf{t})+\alpha(\bar{\mathbf{t}}):
\mathbf{t} + \bar{\mathbf{t}} = \bar{\sigma},\ \mathbf{t}, \bar{\mathbf{t}} \in \mathbb{R}^n \right\},
\ \bar{\sigma} \in \mathbb{R}^n;
\]
where $\bar{\Pi}_{\mathcal{C}}(\mathbf{t})=\tilde{\Pi}_{\mathcal{C}}(\mathbf{t}) - \sum_{i=1}^{n}\delta_{[0,1]}(t_i)$ and
$\alpha(\bar{\mathbf{t}})=-\sum_{i=1}^{n}\delta_{[0,1]}(\bar{t}_i)$, with $\tilde{\Pi}_{\mathcal{C}}(\mathbf{t})=-\infty$ if
$\mathbf{t}\not\in [0,1]^n$ and $\delta_{[0,1]}(\cdot)$ the indicator function of set $[0,1]$ (i.e., $\delta_{[0,1]}(x)=0$
if $x\in [0,1]$ and $\delta_{[0,1]}(x)=+\infty$ otherwise). It is clear that both $\bar{\Pi}_{\mathcal{C}}(\cdot)$ and
$\alpha(\cdot)$ are $M^\natural$-concave over $\mathbb{R}^n$. According to \citet[Section~4.2]{Murota2009},
$\bar{\Pi}(\bar{\sigma})$ is $M^\natural$-concave over $\mathbb{R}^n$. Note that the restriction of $\bar{\Pi}(\bar{\sigma})$
on $[0,1]^n$ is exactly $\tilde{\Pi}(\bar{\sigma})$, which therefore is $M^\natural$-concave over $[0,1]^n$ with straightforward
direct verification according to the definition of $M^\natural$-concavity. We then conclude that $\tilde{\Pi}(\cdot)$ is
submodular over $[0,1]^n$ \citep{Murota_Shioura2004}, which implies that $\Pi^*(\cdot)$, as the restriction of
$\tilde{\Pi}(\cdot)$ to $\{0,1\}^n$, is submodular over $\{0,1\}^n$, as desired.

On the second part of the theorem for the case of the capacity game with function bids,
we assume that, for each supplier $i$, the marginal execution cost $c_i(\cdot)=c_i$ is constant
and the marginal reservation cost $e_i(\cdot)$ is non-decreasing. Let $0\leq c_{1}<\cdots<c_{n}\leq\rho$.
We will show that $\Pi_{\mathcal{C}}(\mathbf{t})$ as defined in (\ref{eqn-supply-chain-profit-n}) is
laminar concave in $\mathbf{t}\in\lbrack0,\bar{d}]^{n}$. With constant $c_{i}(\cdot)$ in (\ref{eq-hbar}) we
have $\bar{h}_{i}(x,\mathbf{t}_{-i})=x+\sum_{j=1}^{i-1}t_{j}$, $x\in
\lbrack0,t_{i}]$. Denote $\bar{\rho}_{i}=\rho-c_{i}$ and define
\[
\rho_{i}(x,\mathbf{t}_{-i}\,|\,d)=\left\{
\begin{array}[c]{ll}%
\bar{\rho}_{i}, & \hbox{if $\bar{h}_i(x, \mathbf{t}_{-i}) \le d$;}\\
0, & \hbox{otherwise.}
\end{array}
\right.
\]
We can write
\[
\sum_{i=1}^{n}\int_{0}^{t_{i}}(\rho-c_{i})\bar{F}(\bar{h}_{i}(x,\mathbf{t}_{-i}))dx
=\mathbb{E}_{D}\left[  \sum_{i=1}^{n}\omega_{i}(\mathbf{t}\,|\,d)\right]  ,
\]
where
\[
\omega_{i}(\mathbf{t}\,|\,d)=\int_{0}^{t_{i}}\rho_{i}(x,\mathbf{t}_{-i}\,|\,d)dx
=\int_{0}^{b_{i}(\mathbf{t}\,|\,d)}\bar{\rho}_{i}\,dx=\bar{\rho}_{i}\,b_{i}(\mathbf{t}\,|\,d),
\]
and $b_{i}(\mathbf{t}\,|\,d)=\min\{(d-\sum_{j=1}^{i-1}t_{j})^{+},t_{i}\}$.
Denote $a_{i}=d-\sum_{j=1}^{i}t_{j}$ for $i=0,1,\ldots,n$. Then
\[
b_{i}(\mathbf{t}\,|\,d)=\min\{a_{i-1}^{+},t_{i}\}=a_{i-1}^{+}-(a_{i-1}^{+}-t_{i})^{+}=a_{i-1}^{+}-a_{i}^{+},
\]
which leads to
\[
\omega_{i}(\mathbf{t}\,|\,d)=\bar{\rho}_{i}\left(  a_{i-1}^{+}-a_{i}^{+}\right).
\]
Therefore,
\begin{align*}
\sum_{i=1}^{n}\omega_{i}(\mathbf{t}\,|\,d)
 &  =\bar{\rho}_{1}\,d-\sum_{i=1}^{n-1}(\bar{\rho}_{i}-\bar{\rho}_{i+1})a_{i}^{+}-\bar{\rho}_{n}a_{n}^{+}\\
&  =\bar{\rho}_{1}\,d-\sum_{i=1}^{n-1}(c_{i+1}-c_{i})\left(  d-\sum_{j=1}^{i}t_{j}\right)^{+}
  -\bar{\rho}_{n}\left(  d-\sum_{j=1}^{n}t_{j}\right)^{+}.
\end{align*}
Noticing that both $(d-\tau)^{+}$ and $\int_{0}^{\tau}e_{i}(x)dx$ ($1\leq i\leq n$) are convex in $\tau\geq0$,
we conclude that $\Pi_{\mathcal{C}}(\mathbf{t})$ is laminar concave with the corresponding laminar family
$\mathcal{L} = \{\{i\}: i\in N\}\cup\{\{1,\ldots, i\}: i\in N\}$.

\end{document}